\documentclass[11pt,3p,times]{elsarticle}

\usepackage{amsmath,amsfonts,amssymb}
\usepackage{amsthm}
\usepackage{bm}
\usepackage{graphicx,epstopdf}
\usepackage[position=top,labelfont=normalfont,textfont=normalfont,singlelinecheck=off,justification=raggedright]{subfig}
\usepackage{floatrow}
\usepackage[dvipsnames]{xcolor}
\usepackage{setspace}
\usepackage{enumerate,etaremune}
\usepackage{booktabs}
\usepackage{tabularx,multirow}
\usepackage{framed}
\usepackage{hhline}
\usepackage{url}
\usepackage[unicode,bookmarks=false]{hyperref}
\hypersetup{
  colorlinks,
  citecolor=Blue,linkcolor=Blue,urlcolor=Blue
}

\usepackage{lineno}

\usepackage[textsize=footnotesize,linecolor=red,backgroundcolor=red!25,bordercolor=red]
  {todonotes}

\usepackage{tikz}
\usetikzlibrary{shapes.geometric}
\usetikzlibrary{shapes,arrows}
\usetikzlibrary{plotmarks}

\usetikzlibrary{calc}

\usepackage{algorithm}
\usepackage{algorithmicx,algpseudocode}
\usepackage{cases}

\newcommand{\eg}{{\it e.g.}}
\newcommand{\ie}{{\it i.e.}}
\newcommand{\cf}{{\it cf.}}
\newcommand{\etal}{{\it et al.}}
\newcommand{\tensor}[1]{\bm{#1}}
\newcommand{\stress}{\sigma}

\newcommand{\tstress}{\tensor{\stress}}

\newcommand{\dd}{\mathrm{d}}
\newcommand{\pd}{\partial}
\newcommand{\Del}{\mathrm{\Delta}}
\newcommand{\el}{\mathrm{e}}
\newcommand{\pl}{\mathrm{p}}

\newcommand{\rn}[1]{\uppercase\expandafter{\romannumeral #1\relax}}

\DeclareMathOperator{\grad}{\nabla}
\DeclareMathOperator{\diver}{\nabla\cdot}

\DeclareMathOperator{\trace}{tr}

\newsavebox{\dotbox}

\theoremstyle{remark}

\definecolor{kaist-blue}{RGB}{20,135,200}
\definecolor{kaist-dark-blue}{RGB}{0,65,145}
\definecolor{kaist-medium-blue}{RGB}{0,65,135}
\definecolor{kaist-light-blue}{RGB}{95,190,235}
\definecolor{kaist-dark-gray}{RGB}{124,124,124}

\newcommand{\revised}[1]{{\color{black} #1}}

\newcolumntype{L}[1]{>{\raggedright\let\newline\\arraybackslash\hspace{0pt}}m{#1}}
\newcolumntype{C}[1]{>{\centering\let\newline\\arraybackslash\hspace{0pt}}m{#1}}
\newcolumntype{R}[1]{>{\raggedleft\let\newline\\arraybackslash\hspace{0pt}}m{#1}}

\allowdisplaybreaks

\biboptions{sort&compress,square,comma,numbers}
\AtBeginDocument{\hypersetup{citecolor=MidnightBlue,linkcolor=MidnightBlue,urlcolor=MidnightBlue}}

\usepackage{etoolbox}
\makeatletter
\patchcmd{\ps@pprintTitle}
{Preprint submitted to}
{~}
{}{}
\makeatother

\begin{document}

\begin{frontmatter}

\title{Hybrid continuum--discrete simulation of granular impact dynamics}

\author[HKU]{Yupeng Jiang}
\author[KAIST]{Yidong Zhao}
\author[HKU]{Clarence E. Choi\corref{corr}}
\ead{cechoi@hku.hk}
\author[KAIST]{Jinhyun Choo\corref{corr}}
\ead{jinhyun.choo@kaist.ac.kr}

\cortext[corr]{Corresponding Authors}

\address[HKU]{Department of Civil Engineering, The University of Hong Kong, Hong Kong}
\address[KAIST]{Department of Civil and Environmental Engineering, KAIST, South Korea}

\journal{~}

\begin{abstract}
Granular impact -- the dynamic intrusion of solid objects into granular media -- is widespread across scientific and engineering applications including geotechnics.
Existing approaches to the simulation of granular impact dynamics have relied on either a purely discrete method or a purely continuum method.
Neither of these methods, however, is deemed optimal from the computational perspective.
Here, we introduce a hybrid continuum--discrete approach, built on the coupled material-point and discrete-element method (MP-DEM), for simulation of granular impact dynamics with unparalleled efficiency.
To accommodate highly complex solid--granular interactions, we enhance the existing MP-DEM formulation with three new ingredients: 
(i) a robust contact algorithm that couples the continuum and discrete parts without any interpenetration under extreme impact loads, 
(ii) large deformation kinematics employing multiplicative elastoplasticity,
and
(iii) a trans-phase constitutive relation capturing gasification of granular media.
For validation, we also generate experimental data through laboratory measurement of the impact dynamics of solid spheres dropped onto dry sand.
Simulation of the experiments shows that the proposed approach can well reproduce granular impact dynamics in terms of impact forces, intrusion depths, and splash patterns.
Further, through parameter studies on material properties, model formulations, and numerical schemes, we identify key factors for successful continuum--discrete simulation of granular impact dynamics.
\end{abstract}


\end{frontmatter}


\section{Introduction}
\label{sec:Intro}
Granular impact -- the dynamic intrusion of solid objects into granular media -- is common in a variety of scientific and engineering problems.
Examples in geotechnical engineering include dynamic compaction and rockfall protection~\cite{mayne1983impact,lee2004method,pichler2005impact,lambert2013design,ng2016large,koo2017dynamic,ng2018comparison,kwan2019finite}.
The physical process in granular impact involves extremely complex and rapid interactions among the intruder and grains, which have attracted a large number of experimental and modeling studies alike (\eg~\cite{pan2002simulation,katsuragi2007unified,umbanhowar2010granular,clark2012particle,clark2014collisional,clark2016steady}).

So far, the discrete element method (DEM) has served as the primary means for the simulation of granular impact dynamics (\eg~\cite{wada2006numerical,teufelsbauer2011simulation,ma2014numerical,shen2019analyses,shen2020discrete,naito2020rockfall,xu2020buffering,su2021effects}).
The key advantage of DEM is that it explicitly incorporates the particulate nature of granular media and their interaction with solid intruders.
Unfortunately, however, the extremely high computational cost of DEM inhibits its use for modeling real-world granular materials (\eg~sand) comprised of a countless number of irregularly-shaped particles.
Accordingly, DEM-based studies have restricted their attention to investigating the fundamental physics of idealized granular materials under the impact of solid intruders.

Meanwhile, Dunatunga and Kamrin~\cite{dunatunga2017continuum} have proposed a fully continuum approach for modeling solid intruders and granular materials, employing the material point method (MPM)~\cite{sulsky1995application,bardenhagen2000material,bardenhagen2004generalized} to accommodate large deformations.
Continuum modeling of granular materials is not only far more efficient than discrete modeling but also sufficiently accurate in many cases provided that proper formulations and parameters are used.
From the computational perspective, however, continuum representation of a solid intruder would still be sub-optimal.
For example, one must solve the governing equations inside the solid intruder although the intruder almost behaves as a rigid body.
Also, the much higher stiffness of the solid intruder gives rise to a computational bottleneck because the maximum time step size in explicit methods is governed by the highest value of material stiffness.
These drawbacks are particularly undesirable for 3D simulations.

A promising alternative to the existing approaches is a hybrid continuum--discrete approach that describes a granular material as a continuum and a solid intruder as a discrete entity (\eg~\cite{jiang2020hybrid,zhan2020sph}).
One of such hybrid schemes is the material-point and discrete-element method (MP-DEM)~\cite{jiang2020hybrid}, whereby MPM and DEM are coupled by exchanging contact force information at their interface.
The MP--DEM approach has a high potential to simulate a variety of solid--granular interactions with far greater robustness and efficiency compared with pure continuum or discrete methods.

However, granular impact dynamics is beyond the capabilities of the existing MP-DEM approach. 
Three critical reasons are as follows.
First, the existing contact algorithm that couples MPM and DEM is not robust enough to address a high force impact.
Specifically, the existing algorithm does not guarantee satisfaction of the non-penetration constraint, which may be detrimental to the accuracy of numerical solutions under extreme impact conditions.
Second, the existing approach relies on linearized kinematics and Jaumann stress rate, which is increasingly erroneous as deformation becomes large.
Last but not least, the existing MP-DEM formulation does not account for gasification of granular media, in which the state of granular media becomes a low-temperature gas.
As shown by Dunatunga and Kamrin~\cite{dunatunga2015continuum,dunatunga2017continuum}, however, capturing gasification is critical to accurate simulation of granular media undergoing extreme deformation.

In this work, we enhance the existing MP-DEM approach to simulate granular impact dynamics and other similar complex solid--granular interactions.
The enhancements are made in the following three ways.
First, we introduce a highly robust contact algorithm, motivated by the recently developed barrier method for frictional contact~\cite{li2020incremental}, to couple the continuum and discrete parts without any interpenetration.
Second, we reformulate the continuum part of MP-DEM based on finite deformation kinematics, employing multiplicative elastoplasciticy~\cite{lee1969elastic} which has recently been shown robust and accurate for modeling granular flow~\cite{favero-neto2018continuum}. 
Third, we extend the trans-phase constitutive relation of Dunatunga and Kamrin~\cite{dunatunga2015continuum} to the hybrid continuum--discrete setting.
Through these approaches, we enable MP-DEM to simulate more challenging solid--granular interactions robustly and accurately, without any compromise in the efficiency of the original method.

To generate data for validation, we also conduct laboratory experiments measuring the impact dynamics of solid spheres dropped onto dry sand. 
The motivation for performing our own experiments is the lack of benchmark experimental data on solid impacts on real-world granular media, such as sands, for which a continuum description is indeed useful.
Specifically, previous experiments on granular impact (\eg~\cite{clark2012particle,clark2014collisional,clark2016steady}) have used manufactured photoelastic discs -- quite different from granular media in engineering practice -- to decipher the fundamental physics of the process.
In this regard, the experimental results presented in this paper may also be useful for other future engineering-oriented studies.

Through simulation of the experiments, we validate the proposed method in terms of the impact forces, intrusion depths, and splash patterns of granular impact.
Further, we conduct parameter studies on material properties, model formulations, and numerical schemes, to identify key factors for successful continuum--discrete simulation of granular impact dynamics.
\revised{At this point, we note that the simulations presented in this paper focus on a single discrete object because the laboratory experiments use a single object, not because the simulation method is limited to a single object. The proposed method, which is an enhanced version of the MP--DEM~\cite{jiang2020hybrid}, can be well applied to simulate interactions between multiple discrete objects and granular continua as the original MP--DEM.}

\section{Continuum modeling of granular media}
In this section, we formulate a continuum model of granular media subjected to solid impact, incorporating large deformation kinematics, multiplicative elastoplasticity, and a trans-phase constitutive relation.
We then discretize the continuum formulation using MPM, using two types of schemes which have different properties for the energy and angular momentum conservation.

\subsection{Large deformation kinematics}
For an accurate description of large-deformation kinematics of granular media in impact dynamics, we use finite strain theory which distinguishes between the reference configuration and the current configuration.
Let $\bm{\varphi}(\bm{X},t)$ denote the motion of the granular continuum, where $\bm{X}$ denotes the position vector in the reference configuration and $t$ denotes the time.
The displacement vector field is then given by $\bm{u}(\bm{X},t):=\bm{\varphi}(\bm{X},t) - \bm{X}$.
The deformation gradient is defined as
\begin{equation}
  \bm{F} := \frac{\pd \bm{\varphi}}{\pd \bm{X}} = \bm{1} + \frac{\pd \bm{u}}{\pd \bm{X}},
  \label{eq:deformation-gradient}
\end{equation}
where $\bm{1}$ is the second-order identity tensor.
It follows the definition of the Jacobian $J:=\det \bm{F}>0$, which maps the differential volume in the reference configuration ($\dd V$) to that in the current configuration ($\dd v$).
Also, the material time derivative of an arbitrary field variable $A$ is defined as
\begin{equation}
    \dot{A} \equiv \frac{\dd A}{\dd t} := \frac{\pd A}{\pd t} + \grad A\cdot\bm{v},
    \label{eq:material-time-derivative}
\end{equation}
where $\grad$ is the gradient operator evaluated with respect to the current configuration, and $\bm{v} := \dot{\bm{u}}$ is the velocity vector.

To distinguish between elastic and plastic deformations, we postulate that the deformation gradient can be decomposed in a multiplicative manner~\cite{lee1969elastic}
\begin{equation}
    \bm{F} = \bm{F}^{\el}\cdot\bm{F}^{\pl}
    \label{eq:multiplicative-decomposition}
\end{equation}
where superscripts $(\cdot)^{\el}$ and $(\cdot)^{\pl}$ denote the elastic and plastic parts, respectively. 
The upshot of this multiplicative decomposition is that it allows one to model large-deformation elastoplasticity without the need to employ a specific type of objective stress rate, which is highly desirable as special care should be exercised to use an objective stress rate correctly (see, \eg~\cite{bazant2012work,ji2013importance}).
We also note that multiplicative elastoplasticity has recently been shown robust and accurate for continuum modeling of granular flow~\cite{favero-neto2018continuum}.

\subsection{Constitutive models}
In what follows, we introduce a set of constitutive formulations for granular media subjected to extremely large deformations.
They will be presented in the following sequence: (i) elasticity, (ii) plasticity, and (iii) a trans-phase relation accounting for gasification.

For modeling the elastic behavior, we use hyperelasticity in which the stress tensor is derived from an explicitly defined strain energy function.
Note that the use of hyperelasticity is required to bypass the use of an objective stress rate~\cite{simo1985unified}.
Under the assumption of isotropic elastic behavior, we express the strain energy density using the elastic part of the left Cauchy--Green deformation tensor, $\bm{b}^{\el} := \bm{F}^{\el}\cdot(\bm{F}^{\el})^{\sf T}$.
The hyperelastic relation is then given by
\begin{equation}
    \tstress = \frac{1}{J}\left(2\bm{b}^{\el}\cdot\frac{\pd \psi(\bm{b}^{\el})}{\pd \bm{b}^{\el}}\right),
    \label{eq:hyperelastic-relation}
\end{equation}
where $\tstress$ is the Cauchy stress tensor, and $\psi$ is the strain energy density function. 
As for the specific elasticity model, we use Hencky elasticity which has been commonly used for modeling large deformation behavior of various materials including granular and porous media~\cite{anand1979hencky,choo2018large,zhao2020stabilized}. The energy density function of Hencky elasticity is given by
\begin{equation}
    \psi(\bm{b}^{\el}) = \frac{1}{2}\lambda(\ln J^{\el})^2 + G \trace \left(\frac{1}{2} \ln\bm{b}^{\el}\right)^{2}.
    \label{eq:energy-density-function}
\end{equation}
where $J^{\el}=\det \bm{F}^{\el}$ is the elastic part of the Jacobian.

For modeling the plastic behavior, in this work we consider two types of constitutive models: (i) Drucker--Prager plasticity~\cite{drucker1952soil}, which is a standard rate-independent plasticity model for sands, and (ii) the $\mu(I)$ rheology~\cite{jop2006constitutive}, which is a popular rate-dependent model for granular flow. 
The motivation for employing these two models is to investigate the role of the rate-dependence of plastic flow in granular impact dynamics, by repeating the same simulation with the two different constitutive models.
We note that the $\mu(I)$ model can be interpreted as an extension of the Drucker--Prager model to rate-dependent plastic flow, as described below.

For cohesionless materials, the Drucker--Prager and $\mu(I)$ models share the common form of the yield function, which can be written as
\begin{equation}
    \mathcal{F}(p,\tau) = \tau - \bar{\mu} p \leq 0,
    \label{eq:yield-unified}
\end{equation}
where $p := -(1/3)\trace(\tstress)$ is the mean normal pressure, and $\tau:=\sqrt{1/2}\|\tstress - p\bm{1}\|$, and 
$\bar{\mu}$ is a coefficient related to the frictional resistance of the granular material.
When the plastic flow of the Drucker--Prager model is assumed to be isochoric, the two models also use the same flow rule whereby the the plastic strain is purely deviatoric.
Therefore, the $\mu(I)$ rheology model can be utilized without significant change in the existing algorithm for the Drucker--Prager plasticity model~\cite{borja2013plasticity,klar2016drucker}.

The Drucker--Prager and $\mu(I)$ models are distinguished according to the specific forms and rate dependence of $\bar{\mu}$, as
\begin{equation}
    \bar{\mu} = 
    \begin{cases}
    \mu_{s} & \text{Drucker--Prager (rate independent)},\\
    \mu(I) & \text{$\mu(I)$ (rate dependent)}.
    \end{cases}
\end{equation}
When the Drucker--Prager model is fitted to the compression corners of the Mohr--Coulomb failure surface, $\mu_{s}$ is related to the friction angle $\phi$ as~\cite{borja2013plasticity}
\begin{equation}
    \mu_{s} = \dfrac{2\sqrt{3}\sin\phi}{3 - \sin\phi}.
    \label{eq:stress-ratio-friction-angle}
\end{equation}
The friction angle in Drucker--Prager plasticity is usually considered constant and hence rate independent.
In the $\mu(I)$ rheology, however, the frictional resistance of a granular flow does depend on the rate of the magnitude of plastic shear strain, $\dot{\gamma}^p$. 
Specifically, the $\mu(I)$ rheology defines the dimensionless inertial number of a granular flow, $I := \dot{\gamma}^p \sqrt{d^2 \rho_s/p}$ ($d$ is the mean particle size and $\rho_s$ is the particle density) and relate $\mu$ and $I$ as
\begin{equation}
    \mu(I) = 
    \begin{cases}
        \mu_s & \text{if}\;\; I = 0\,, \\
        \mu_s + \dfrac{\mu_2 - \mu_s}{I_0/I + 1} & \text{if}\;\; I > 0,
    \end{cases}
    \label{eq:muI-model}
\end{equation}
where $I_0$ is a constant material parameter, and $\mu_2$ is the upper limit of the frictional resistance during plastic flow, which may be linked to a friction angle as in Eq.~\eqref{eq:stress-ratio-friction-angle}.

Lastly, to capture gasification of granular media which may occur when grains are detached from each other, we employ the so-called trans-phase constitutive relation proposed by Dunatunga and Kamrin~\cite{dunatunga2015continuum}.
In essence, the trans-phase relation treats the bulk modulus of a granular material, $K$, as a function of the mass density, $\rho$, based on an equation of state.
Specifically,
\begin{equation}
    K(\rho) = 
    \begin{cases}
        0 & \text{if} \;\; \rho<\rho_{c}, \\
        K_c & \text{if} \;\; \rho\geq \rho_{c},
    \label{eq:trans-phaseK}
    \end{cases}
\end{equation}
where $K_c$ is the bulk modulus when the grains are considered in contact, and $\rho_{c}$ is the critical value of density below which the grains are gasified. 
This trans-phase relation has proven capable of modeling granular gasification in a wide range of granular flow and solid--granular interaction problems~\cite{dunatunga2015continuum,dunatunga2017continuum}.

\subsection{MPM discretization}
We now discretize the continuum formulation for granular media using MPM.
In a nutshell, MPM traces state variables (\eg~mass and velocity) at continuum particles (material points) and solves the governing equation in a background grid that exchanges information with the particles.
In this way, MPM can effectively handle large deformations without the issue of mesh distortion.
One may also view MPM as a variant of the finite element method (FEM) in which the quadrature points are detached from the grid and movable.

As MPM is an updated Lagrangian approach, we write the governing equation -- linear momentum balance -- in the current configuration as
\begin{equation}
    \rho\dot{\bm{v}} = \diver\bm{\sigma} + \rho\bm{g},
    \label{eq:linear-momentum-balance}
\end{equation}
where $\diver$ is the divergence operator evaluated with respect to the current configuration, 
and $\bm{g}$ is the gravitational force vector.
The MPM discretization starts with the same Galerkin procedure as in the finite element method (FEM).
Denoting nodal quantities in a background grid by subscript $(\cdot)_{i}$, the Gakerlin method discretizes Eq.~\eqref{eq:linear-momentum-balance}  as
\begin{equation}
    m_{i}\dot{\bm{v}}_{i}=\bm{f}_{i},
    \label{eq:grid-momentum}
\end{equation}
where $m_{i}$ is the mass, $\bm{v}_{i}$ is the velocity vector, and $\bm{f}_{i}$ is the force vector which is the sum of the internal and external force vectors. 
While Eq.~\eqref{eq:grid-momentum} has been obtained in the same way as FEM, the succeeding procedure of MPM is different in that the particles -- corresponding to the quadrature points in FEM -- are not fixed to the grid.

\begin{figure}[h!]
    \centering
    \includegraphics[width = 1.0\textwidth]{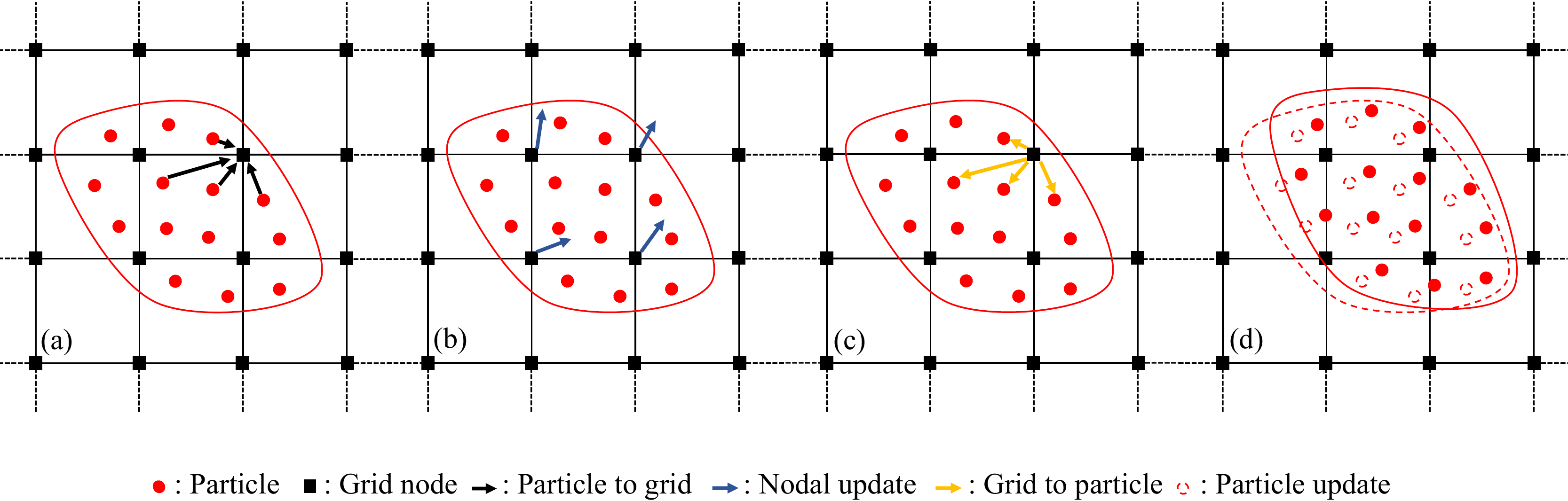}
    \caption{MPM procedure: (a) particle-to-grid (P2G) transfer, (b) nodal update, (c) grid-to-particle (G2P) transfer, (d) particle update.}
    \label{fig:MPM-procedure}
\end{figure}

The MPM procedure in each time step -- illustrated in Fig.~\ref{fig:MPM-procedure} -- can be recapitulated as follows.
In the beginning, the quantities of particles are mapped to nodes in the background grid, which is often called the particle-to-grid (P2G) transfer.
The P2G transfer of linear momentum is critical and deserves elaboration.
Let us use subscript $(\cdot)_{p}$ to denote particle-wise quantities and denote quantities at the current time step without any indices for simplicity.
The general affine momentum transfer scheme can be written as~\cite{jiang2015affine,jiang2017angular}
\begin{equation}
    m_{i}\bm{v}_i = \sum_p w_{ip} m_p[\bm{v}_{p} + \bm{B}_{p}(\bm{D}_{p})^{-1}(\bm{x}_{i}-\bm{x}_{p})].
    \label{eq:P2G-momentum}
\end{equation}
Here, $w_{ip}$ is the interpolation weights, 
$\bm{B}_{p}$ is a matrix containing the angular momentum information,
$\bm{D}_{p}$ is the inertia matrix for affine motion, 
and $\bm{x}_{i}$ and $\bm{x}_{p}$ are the position vectors of the nodes and the particles, respectively.
It is noted that setting $\bm{B}_{p}=\bm{0}$ leads to the classic non-affine transfer scheme in MPM, which may not conserve the angular momentum.
Following the P2G transfer, the nodal momentum is updated through an explicit time integration
\begin{equation}
    (m_{i}\bm{v}_{i})^{n+1} = m_{i}\bm{v}_{i} + \Delta{t} {\bm{f}_{i}},
    \label{eq:grid-update}
\end{equation}
where superscript $(\cdot)^{n+1}$ denotes the updated quantities, and $\Delta{t}$ is the time increment.
The updated nodal quantities in the grid are then mapped back to particles, which is usually referred to as the grid-to-particle (G2P) transfer.
A critical consideration at this stage is the velocity update scheme -- the way how the particle velocity is updated based on nodal velocities -- as it controls the numerical damping and stability. 
In general, the velocity scheme is a combination of the fluid-implicit-particle (FLIP) method~\cite{brackbill1986flip}, whereby the increment of the particle velocity is interpolated from nodes, and the particle-in-cell (PIC) method~\cite{harlow1964particle}, whereby the total particle velocity is interpolated from nodes.
So a general form of the velocity scheme can be written by introducing the blending ratio of FLIP and PIC, $\eta\in[0,1]$, as:
\begin{equation}
    \bm{v}_{p}^{n+1} = 
    \eta\underbrace{\left(\bm{v}_{p} + \sum_{i}w_{ip}(\bm{v}_{i}^{n+1}-\bm{v}_{i})\right)}_\text{FLIP}
    + (1 - \eta)\underbrace{\left(\sum_{i}w_{ip}\bm{v}_{i}^{n+1}\right)}_{\text{PIC}}\,.
    \label{eq:G2P-velocity}
\end{equation}
This blending is motivated by that FLIP has much less dissipative than PIC but is more unstable.
Lastly, the particles positions are updated as
\begin{equation}
    \bm{x}_{p}^{n+1}=\bm{x}_{p} + \Delta{t} \sum_{i}w_{ip}\bm{v}_{i}^{n+1}.
\end{equation}
If desired, the background grid may be reset before proceeding to the next step.

As mentioned above, a few different schemes are available for the momentum transfer and the velocity update, and they have significant impacts on the simulation result.
Specifically, while all the standard MPM schemes conserve mass and linear momentum, the conservation properties of energy and angular momentum are different depending on the transfer/update schemes.
For example, the most common scheme in MPM is FLIP (with a little blending with PIC for stability) for the velocity update, along with the non-affine P2G transfer.
This scheme, however, does not conserve the angular momentum whenever the mass matrix is lumped for a fully explicit time integration.
To address this drawback, Jiang \etal~\cite{jiang2015affine,jiang2017angular} have developed the affine PIC (APIC) method, which conserves the angular momentum even when a lumped mass matrix is used.
Nevertheless, APIC may be too dissipative to simulate granular impact dynamics as the method is rooted on PIC.
Unfortunately, none of the existing transfer/update schemes can perfectly conserve both the energy and angular momentum, so one needs to choose a specific scheme considering the nature of the problem at hand.
However, it remains unknown as to how the choice of the transfer/update scheme affects MPM simulation results of granular impact dynamics.

To examine the effects of the transfer/update scheme on granular impact simulation, here we implement two types of schemes: FLIP and APIC. 
For FLIP, we set $\bm{B}_{p}=\bm{0}$ in Eq.~\eqref{eq:P2G-momentum}, use the interpolation functions of the generalized interpolation material point method (GIMP)~\cite{bardenhagen2004generalized}, and assign $\eta=0.995$ in Eq.~\eqref{eq:G2P-velocity}.
Note that $\eta$ is slightly less than 1 to provide a little amount of damping for stability.
For APIC, we use
\begin{equation}
    \bm{B}_{p} = \sum_{i} w_{ip} \bm{v}_{i}(\bm{x}_{i} - \bm{x}_{p})^{\sf T}
\end{equation}
for Eq.~\eqref{eq:P2G-momentum}, which provably conserves the angular momentum~\cite{jiang2017angular}.
Also, following the original APIC~\cite{jiang2015affine,jiang2017angular}, we employ cubic B-splines for the interpolation functions and the pure PIC method for the velocity update, \ie~set $\eta=0$ in Eq.~\eqref{eq:G2P-velocity}.

\section{Coupling of granular continuum with discrete solid}
\label{sec:coupling}

In this section, we couple the MPM-discretized granular continuum with a solid intruder modeled by DEM.
We first link MPM and DEM through their contact force, incorporating gasification based on the trans-phase constitutive relation.
Then, for robust and efficient calculation of the contact force under high force impact, we devise a new contact algorithm that is simple and interpenetration-free.

\subsection{Coupling MPM and DEM incorporating gasification}
\begin{figure}[h!]
    \centering
    \includegraphics[width=0.5\textwidth]{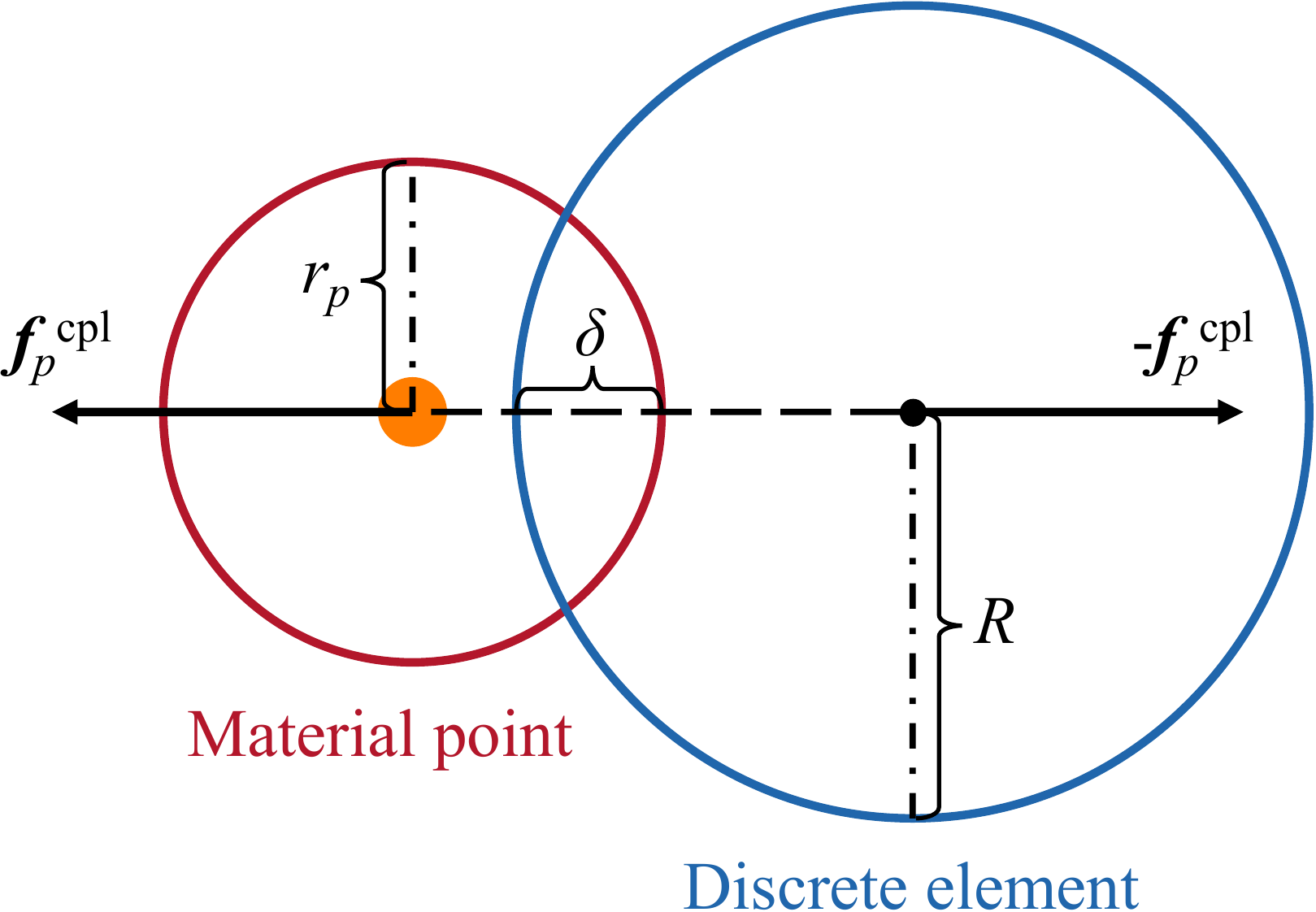}
    \caption{General scheme for coupling MPM and DEM through their contact force.}
    \label{fig:contact-scheme}
\end{figure}

The coupling of MPM (granular continuum) and DEM (solid intruder) begins by calculating the coupling force, which arises from the contact between the continuum and the discrete object, as follows.
When a material point and a discrete element are sufficiently close, the material point is given a radius for contact detection, $r_p$, as illustrated in Fig.~\ref{fig:contact-scheme}.
Then the coupling contact force can be written as $\bm{f}^{\rm cpl}=f\bm{n}$, where $f$ is the force magnitude which is a function of the overlapping distance $\delta$, and $\bm{n}$ is the force direction vector which is parallel to a line connecting the centroids of the material particle and the discrete element.
It is noted that this way to calculate the contact force is the same as that in DEM, except a material point is used for one of the particles. 

The coupling force is then augmented to each of the MPM and DEM calculations.
In doing so, we modify the original MP--DEM~\cite{jiang2020hybrid} to account for gasification of the MPM-discretized continuum.
Recall that the trans-phase constitutive relation~\eqref{eq:trans-phaseK} uses the mass density to detect gasification.
The mass density of a material point is calculated as
\begin{equation}
    \rho_p = \frac{m_{p}}{V_p},
    \label{eq:density-change}
\end{equation}
where $V_p$ is the volume of the material point.
Note that while $m_p$ is constant throughout deformation as the mass is conserved in MPM, $V_p$ is evolving and can be computed by time-integration of the divergence of the velocity field.
According to the value of $\rho_{p}$, the coupling force of individual particles are transferred to the grid (during the P2G stage) as
\begin{equation}
    \bm{f}_{i}^{\rm cpl} 
    = \begin{cases}
        0,& \text{if} \;\; \rho_p<\rho_{c}, \\
        \sum_{p} w_{ip}\bm{f}_{p}^{\rm cpl} & \text{if} \;\; \rho_p\geq \rho_{c}.
    \end{cases}
    \label{eq:contact-force-transfer-trans-phase}
\end{equation}
Following the original trans-phase approaches~\cite{dunatunga2015continuum,dunatunga2017continuum}, $\rho_c$ is set as the mass density in the initial state.
In this way, a gasified material point keeps its coupling contact force locally while having no interaction with its surroundings. 
The grid momentum update equation~\eqref{eq:grid-update} is then modified as
\begin{equation}
    (m_{i}\bm{v}_{i})^{n+1} = m_{i}\bm{v}_{i} + \Del t\left({\bm{f}_{i}} + \bm{f}_{i}^{\rm cpl}\right).
\end{equation}
For DEM, the coupling force $-\bm{f}_p^{\rm cpl}$ is directly added to the resultant force of a discrete element when updating its velocity and position.

\subsection{Barrier method for contact between MPM and DEM}
The existing MP--DEM framework~\cite{jiang2020hybrid} calculates the contact force magnitude using a linear spring model $f=k_{N}\delta$, where $k_{N}$ is the contact normal stiffness.
However, the linear contact model does not guarantee satisfaction of the non-penetration condition, $\delta \leq r_p$, because it cannot prevent the force magnitude from becoming greater than $k_{N}r_p$.
While this limitation may be tolerable for relatively moderate solid--granular interaction problems, it would be a critical drawback for granular impact problems where high contact force is inevitable.

In this work, we propose a new approach -- the barrier method -- for rigorous treatment of the contact between MPM and DEM under extreme solid--granular interactions.
The barrier method has originally been proposed by Li \etal~\cite{li2020incremental} for continuum elastodynamics and then extended to other types of contact problems such as embedded interfaces~\cite{li2021codimensional,lan2021medial,zhao2022barrier}.
These studies have commonly shown that the barrier method is a robust and efficient treatment of challenging contact problems.
Remarkably, the barrier method ensures satisfaction of the non-penetration constraint by construction, and it can be utilized as efficiently as the classic penalty (spring) method.
All the existing barrier contact methods, however, have been formulated for pure continuum problems solved by implicit methods.
Because the problem at hand involves discrete elements and uses an explicit method, we need to modify several aspects of the barrier method.

To formulate a barrier method for contact between a material point and a discrete element, we first define the residual distance as (\cf~Fig.~\ref{fig:contact-scheme})
\begin{equation}
    \xi := r_p - \delta.
\end{equation}
The non-penetration constraint can then be phrased as $\xi \geq 0$.
We note that interpenetration is defined as $\xi < 0$, not $-\delta < 0$, because $r_p$ is the radius of a material point's influence zone -- a numerical parameter -- introduced for MP--DEM coupling rather than the radius of a physical particle.

\begin{figure}[h!]
    \centering
    \includegraphics[width=0.5\textwidth]{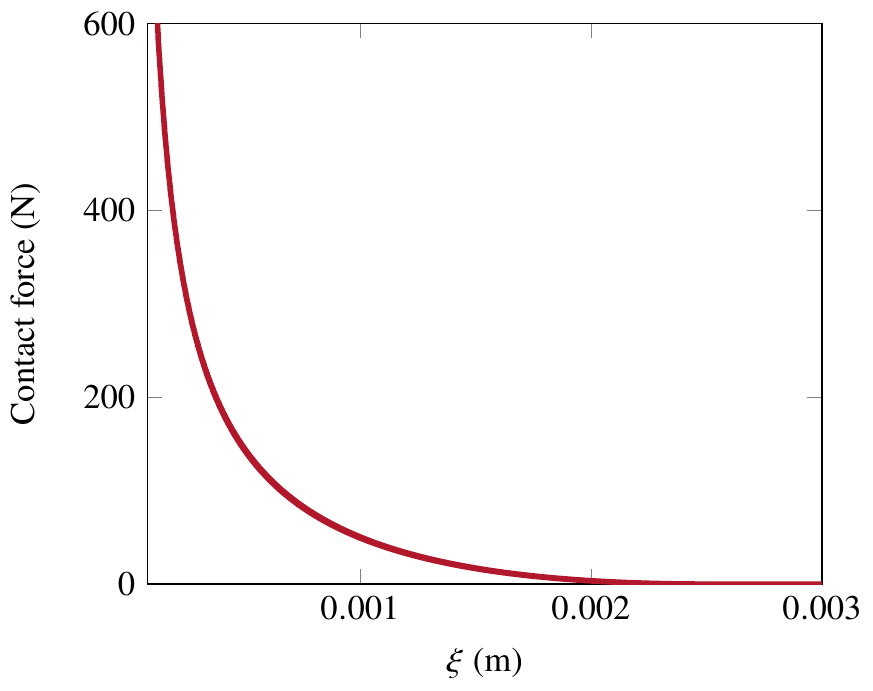}
    \caption{Variation of the contact force with the residual distance ($r_p = 0.0025$ m and $\kappa=10^{4}$ J/m$^2$).}
    \label{fig:barrier-contact-force}
\end{figure}

Next, we introduce a circular elastic barrier whose center coincides with the location of the material point and thickness is equal to $r_p$.
When the barrier is compressed, it stores a potential energy which may be referred to as the barrier energy.
Naturally, the barrier energy must be a function of $\xi$.
Here, we represent the barrier energy function $B(\xi)$ by adapting a $C^{2}$-continuous barrier function from Li \etal~\cite{li2020incremental} to the current problem, as
\begin{equation}
    B(\xi) := 
    \begin{cases}
      -\kappa(\xi-r_p)^2 \ln \left(\dfrac{\xi}{r_p}\right) &\text{if}\;\; 0<\xi\leq r_p, \\
      0 &\text{if}\;\; \xi>r_p,
    \end{cases}
    \label{eq:barrier-energy}
\end{equation}
where $\kappa>0$ is a scalar parameter controlling the stiffness of the barrier, which will be determined shortly.
Note that unlike the original barrier function in Li \etal~\cite{li2020incremental}, Eq.~\eqref{eq:barrier-energy} does not have the free parameter for accuracy control (denoted by $\hat{d}$ therein), because the barrier thickness has been prescribed to be $r_p$ to be consistent with MP--DEM~\cite{jiang2020hybrid}.
Then, based on thermodynamic conjugacy, the contact force magnitude $f$ is derived from the barrier function as
\begin{equation}
    f(\xi) := -\dfrac{\pd {B}(\xi)}{\pd {\xi}}
    =
    \begin{cases}
    \kappa(\xi - r_p)\left[2\ln\left(\dfrac{\xi}{r_p}\right)-\dfrac{r_p}{\xi}+1\right] &\text{if}\;\; 0<\xi < r_p, \\
    0, &\text{if} \;\; \xi \geq r_p.
    \end{cases}
    \label{eq:barrier-contact-force}
\end{equation}
Figure~\ref{fig:barrier-contact-force} shows an example of how the contact force varies with the residual distance.
From the figure as well as from Eq.~\eqref{eq:barrier-contact-force}, one can see that $f>0$ when $\xi < r_p$ and $f\rightarrow \infty$ as $\xi\rightarrow 0$. 
This variation of the contact force ensures that the non-penetration constraint is satisfied under any finite amount of load.

Importantly, when the barrier method is used in conjunction with an explicit method -- as in the present work -- the barrier stiffness parameter $\kappa$ should be determined differently from the original barrier method formulated for an implicit method.
Specifically, in the original method, $\kappa$ is determined such that it yields the best possible condition of the matrix in an implicit solution step.
However, because an explicit method does not solve any matrix problem, a different strategy is necessary to determine the value of $\kappa$.

Here we develop a way to calculate the value of $\kappa$ for an explicit method based on numerical stability. 
For the current MP--DEM, the stability condition is given by:
\begin{equation}
    \Delta t\leq\alpha\Delta t_{\text{crit}},\quad
    \Delta t_{\text{crit}}
    := \min 
    \begin{cases}
      \dfrac{h}{c} \\
      2\pi\sqrt{\dfrac{m_p}{k_N}}
    \end{cases}
    \label{eq:time-step}
\end{equation}
where $\Delta{t}_{\text{crit}}$ is the critical time step size, 
$\alpha \in [0,1]$ is a safety factor for stability, 
$h$ is the grid spacing, 
and $c$ is the acoustic velocity of the continuum.
We note that unlike the stability condition in the original MP--DEM~\cite{jiang2020hybrid}, Eq.~\eqref{eq:time-step} uses the mass of material point $m_p$ to account for potential gasification which makes $m_p$ very small.
When the barrier method is used, $k_N$ is not a constant and given by 
\begin{equation}
    k_N 
    = -\dfrac{\partial{f}(\xi)}{\partial \xi}
    = \begin{cases}
    -\kappa \left[2\ln\left(\dfrac{\xi}{r_p}\right) + \dfrac{(\xi-r_p)(3\xi+r_p)}{\xi^2}\right] &\text{if}\;\; 0<\xi < r_p, \\
    0 &\text{if}\;\; \xi\geq r_p.
    \end{cases}
    \label{eq:barrier-stiffness}
\end{equation}
As can be seen, $k_{N}$ is proportional to $\kappa$.
Ideally, we want $\kappa$ to make $2\pi\sqrt{m_p/k_N} \geq h/c$ such that $\Delta{t}_{\text{crit}}=h/c$ regardless of $\kappa$.
Rearranging Eq.~\eqref{eq:time-step}, we find that this is equivalent to make $k_N \leq k_{N,\text{crit}}$ where
\begin{equation}
    k_{N,\text{crit}} := \dfrac{4m_p\pi^2c^2}{h^2}.
    \label{eq:critical-stiffness}
\end{equation}
Unfortunately, however, $k_N \leq k_{N,\text{crit}}$ is not always possible because $k_N\rightarrow \infty$ as $\xi\rightarrow 0$.
Therefore, we instead find the value of $\kappa$ that makes $k_N \leq k_{N,\text{crit}}$ until $\xi$ is greater than a user-defined parameter $\bar{\xi}$ satisfying $0<\bar{\xi}<r_p$.
Combining Eqs.~\eqref{eq:barrier-stiffness} and~\eqref{eq:critical-stiffness} and inserting $\bar{\xi}$ into it, we get
\begin{equation}
    \kappa = -k_{N,\text{crit}} \left[2\ln\left(\dfrac{\bar{\xi}}{r_p}\right) + \dfrac{(\bar{\xi}-r_p)(3\bar{\xi}+r_p)}{\bar{\xi}^2}\right]^{-1}.
    \label{eq:kappa-calculation}
\end{equation}
In this work, we set $\bar{\xi}=0.2r_p$. 
It is believed that this type of stability-based approach can be used for determining the value of $\kappa$ in general explicit methods.

\revised{Before validating the proposed method, it is noted that all the numerical parameters for coupling MPM and DEM can be calculated from $V_p$, which in turn is a function of the grid spacing and the number of material points per cell. 
Firstly, following the original MP--DEM formulation~\cite{jiang2020hybrid}, the radius of contact detection can be calculated as $r_p=(V_p)^{1/\rm{dim}}$. Once $r_p$ is determined, $\kappa$ is calculated according to Eq.~\eqref{eq:kappa-calculation}.}


\section{Validation}
\label{sec:validation}
In this section, we validate the hybrid continuum--discrete approach with experimental data.
We first conduct laboratory experiments that measure the impact dynamics of solid spheres dropped onto dry sand. 
We then simulate the experiments with the proposed approach and compare the simulation and experimental results quantitatively and qualitatively.

\subsection{Laboratory experiments}
\begin{figure}[h!]
    \centering
    \includegraphics[width=0.5\textwidth]{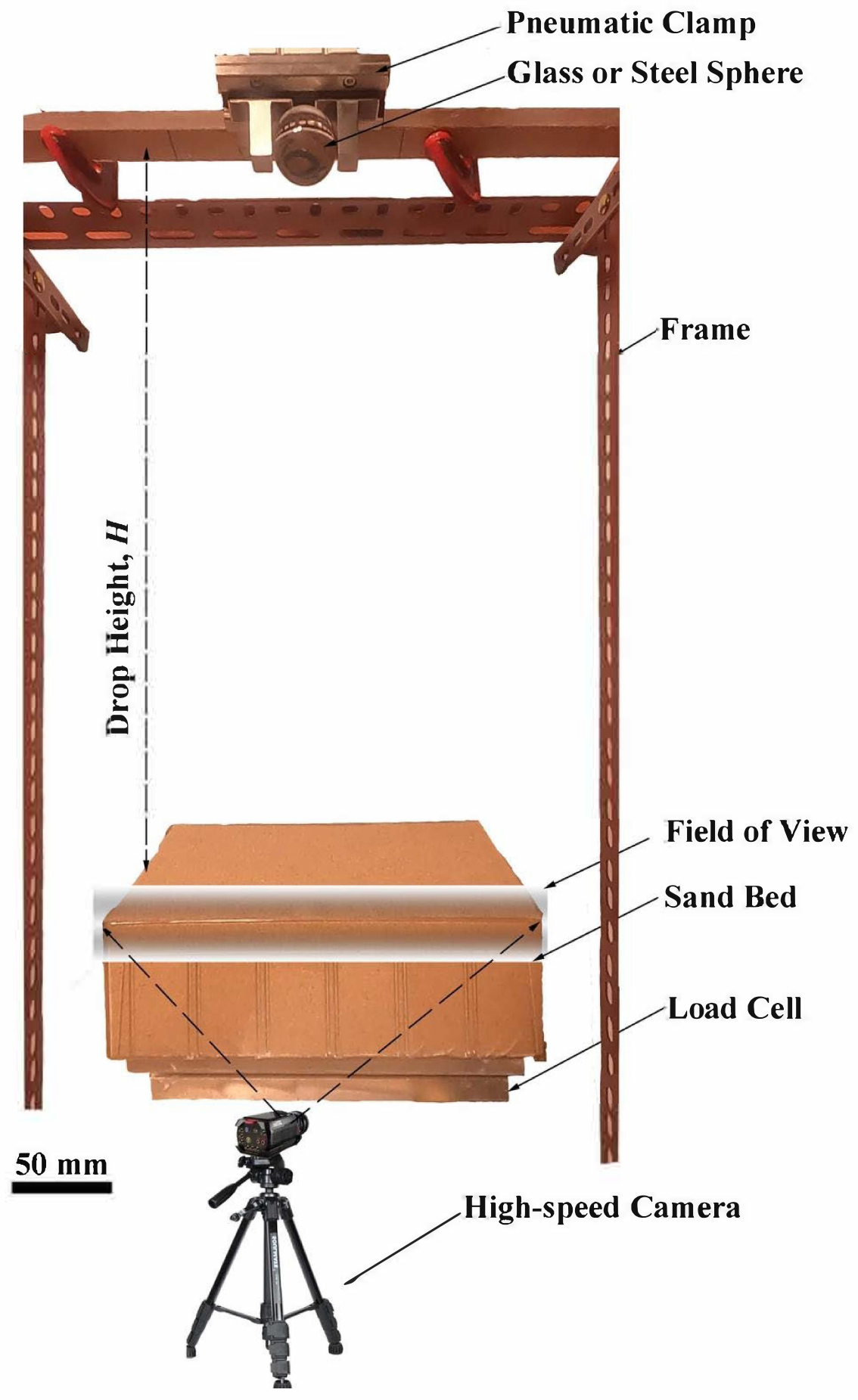}
    \caption{Laboratory test setup.}
    \label{fig:lab-setup}
\end{figure}

Figure~\ref{fig:lab-setup} shows the setup of our laboratory experiments measuring solid impacts on dry sand.
The sand bed was prepared by filling a 22.5 cm wide, 10 cm long, and 8 cm high box with Toyura sand in a fairly loose manner (the dry density was about 1.28 g/cm$^3$).
For the solid intruder, we used two types of spheres: (i) a glass sphere, whose density is 2.29 g/cm$^3$ and diameter is 38.34 mm, and (ii) a steel sphere, whose density was 7.78 g/cm$^{3}$ and diameter was 40.00 mm.
Each sphere was dropped from \revised{three different heights, namely, 20 cm, 40 cm, and 60 cm, onto the center of the sand bed.} 
We measured the impact load through a load cell (frequency: 2000 Hz) installed at the bottom of the sand box.
Using a high-speed (2000 frames per second) camera, we also took snapshots of the granular impact process to estimate intrusion depths and capture splash patterns. 
\revised{Based on the locations of the top of the spheres in the snapshots, we calculated the intrusion depths of the spheres over time. 
We also measured the final depths from the final positions of the spheres.}
\revised{For each experimental setup, we repeated the test six times to ensure reproducibility and to characterize variations of the measured quantities.}

\subsection{Simulation setup}
\begin{figure}[h!]
    \centering
    \includegraphics[width=0.7\textwidth]{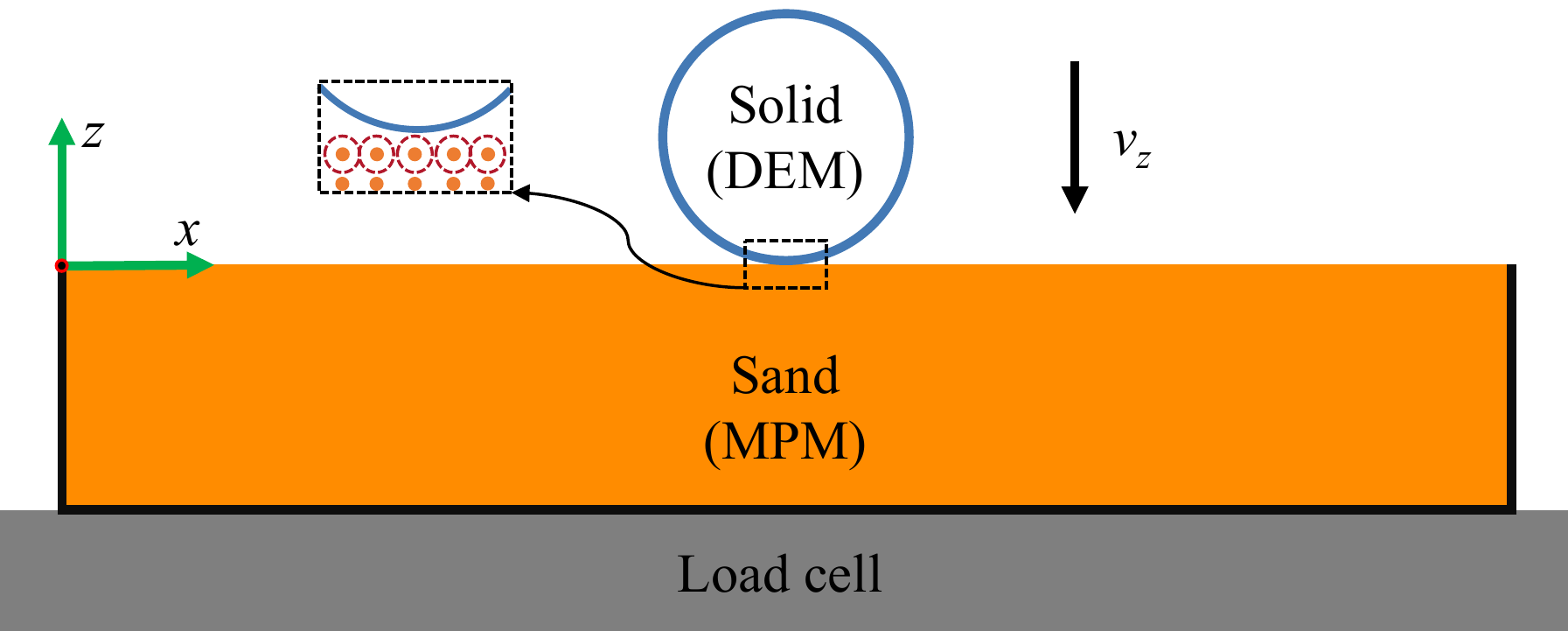}
    \caption{Simulation setup: a section view. Note that the actual simulation is in 3D.}
    \label{fig:simulation-setup}
\end{figure}

We conduct three-dimensional (3D) simulation of the laboratory experiment based on the setup illustrated in Fig.~\ref{fig:simulation-setup}.
For MPM modeling of the sand, we introduce a structured background grid composed of mono-sized ($h = 4$ mm) square cells and implement rigid boundary grids at the bottom and sidewalls to emulate the confined boundary conditions.  
We then discretize the domain by material points using the Poisson disc sampling method~\cite{gamito2009accurate}, which results in approximately four material points per cell.  
The material volume per point ($V_p$) is calculated as the total domain volume divided by the number of material points. 
Next, as for DEM modeling of the solid sphere, we place the sphere right above the sand box and set its initial vertical velocity ($v_z$) equal to the analytical impact velocity. 
Finally, \revised{from $h=4$ mm, the numerical parameters for coupling MPM and DEM are calculated as follows: the coupling radius $r_p=(V_p)^{1/3}=2.5$ mm and the barrier stiffness $\kappa=10^{4}$ N/m.}

We use material parameters measured from our laboratory experiment, except for the mechanical parameters of the sand.
While the mechanical parameters of Toyura sand under moderate to high confining pressures are well known  (\eg~\cite{verdugo1996steady,choo2018mohr}), their values under a very low confining pressure -- of interest in this work -- are not readily available and difficult to measure directly. 
\revised{The most accurate way to estimate the parameters would be to perform back analysis of the specific laboratory experiments.
In most engineering practices, however, one has to predict impact dynamics that has not occured yet, and so data for back analysis are often unavailable. 
Therefore, instead of conducting back analysis of the experimental data, we use a set of rule-of-thumb parameters in soil mechanics that can reproduce the experimental results reasonably well.} 
They are: a Young's modulus of $E=1$ MPa, Poisson's ratio of $\nu=0.2$, and a friction angle of $\phi=30^{\circ}$.
Note that the friction angle is related to $\mu_s$ as in Eq.~\eqref{eq:stress-ratio-friction-angle}.
Later, we will also show how these material parameters control granular impact dynamics.

The simulation proceeds as follows. 
Since the yield strength is pressure dependent, we first initialize the stresses of the material points through a gravitational preloading stage.
The preloading stage is finished when the kinetic energies of all the material points become nearly zero.
Subsequently, we proceed to the granular impact stage, in which the intrusion of the solid sphere is simulated with a constant time increment of $\Delta{t}=2.0\times{10}^{-5}$ s calculated according to Eq.~\eqref{eq:time-step}.
By default, we use FLIP for velocity update.
To measure the impact load, we \revised{monitor the grid force at the bottom grids like the load cell in the experiments} and process it with a standard moving-average filter.

\begin{figure}[h!]
    \centering
    \subfloat[Impact load]{\includegraphics[width=0.5\textwidth]{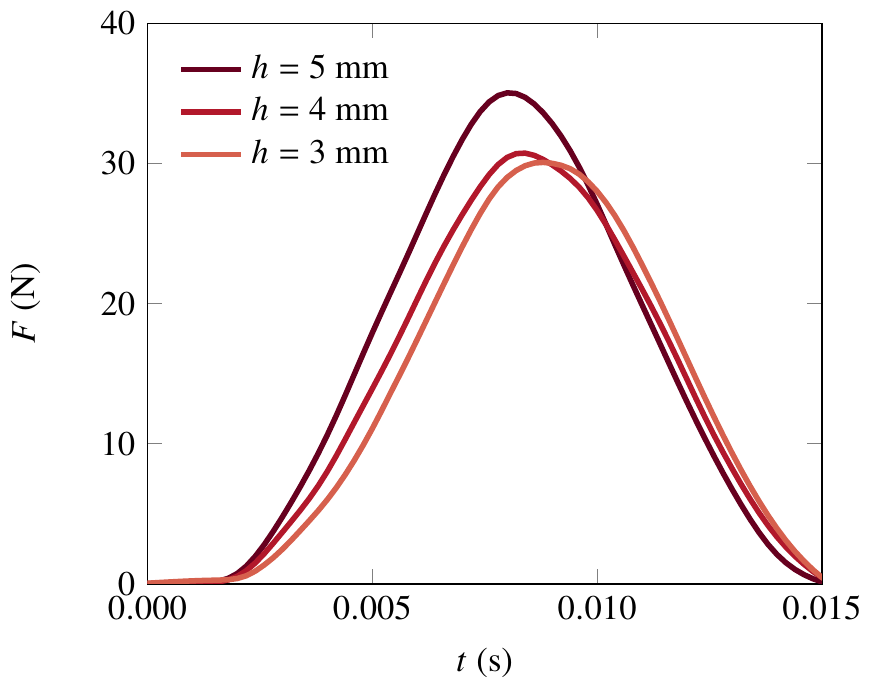}}
    \subfloat[Intrusion Depth]{\includegraphics[width=0.5\textwidth]{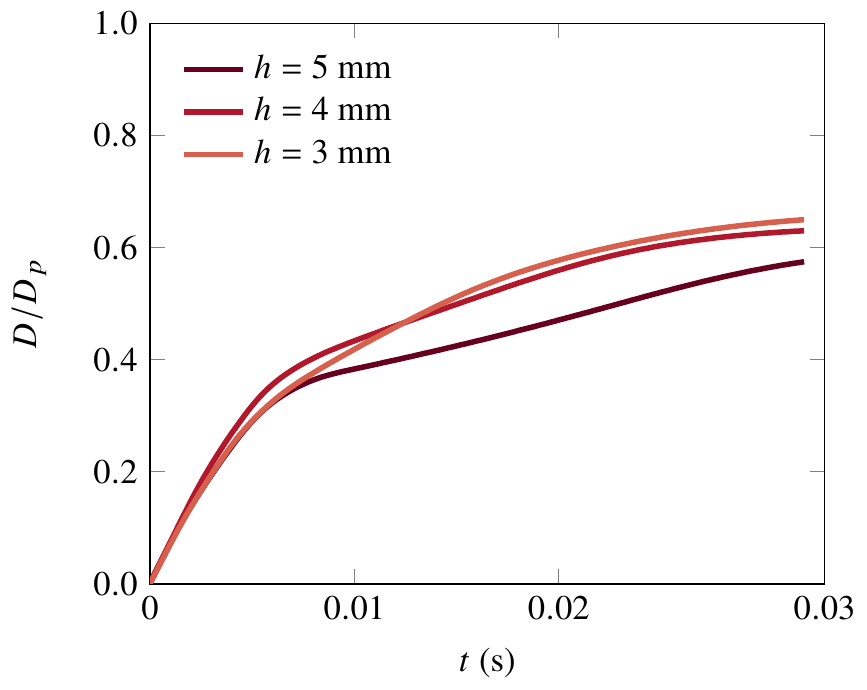}}
    \caption{\revised{Sensitivity of simulation results to the grid size $h$: glass sphere dropped from $H=40$ cm.}}
    \label{fig:mesh-sensitivity}
\end{figure}

\revised{Prior to comparing the experimental and simulation results, we check the sensitivity of the simulation results to the numerical parameters. As explained earlier, all the numerical parameters of the current method are functions of the grid spacing, $h$, and the number of material points per cell. 
Therefore, we examine how the simulation results become different by $h$, keeping the number of material points per cell as four.
In Fig.~\ref{fig:mesh-sensitivity} we show the simulation results for the glass sphere dropped from $H = 40$ cm obtained from three different values of $h$, namely, $3$ mm, $4$ mm, and $5$ mm.
It can be confirmed that the impact load and the intrusion depths -- the quantities of our interest -- show converging trends as $h$ becomes smaller.
Also, $h=4$ mm is deemed sufficiently small in that the results from $h=4$ mm and $h=3$ mm show little difference from a practical viewpoint.
As such, we use $h=4$ mm in the following simulations.}

\subsection{Comparison of experimental and simulation results}
\begin{figure}[h!]
    \centering
    \includegraphics[width=1.0\textwidth]{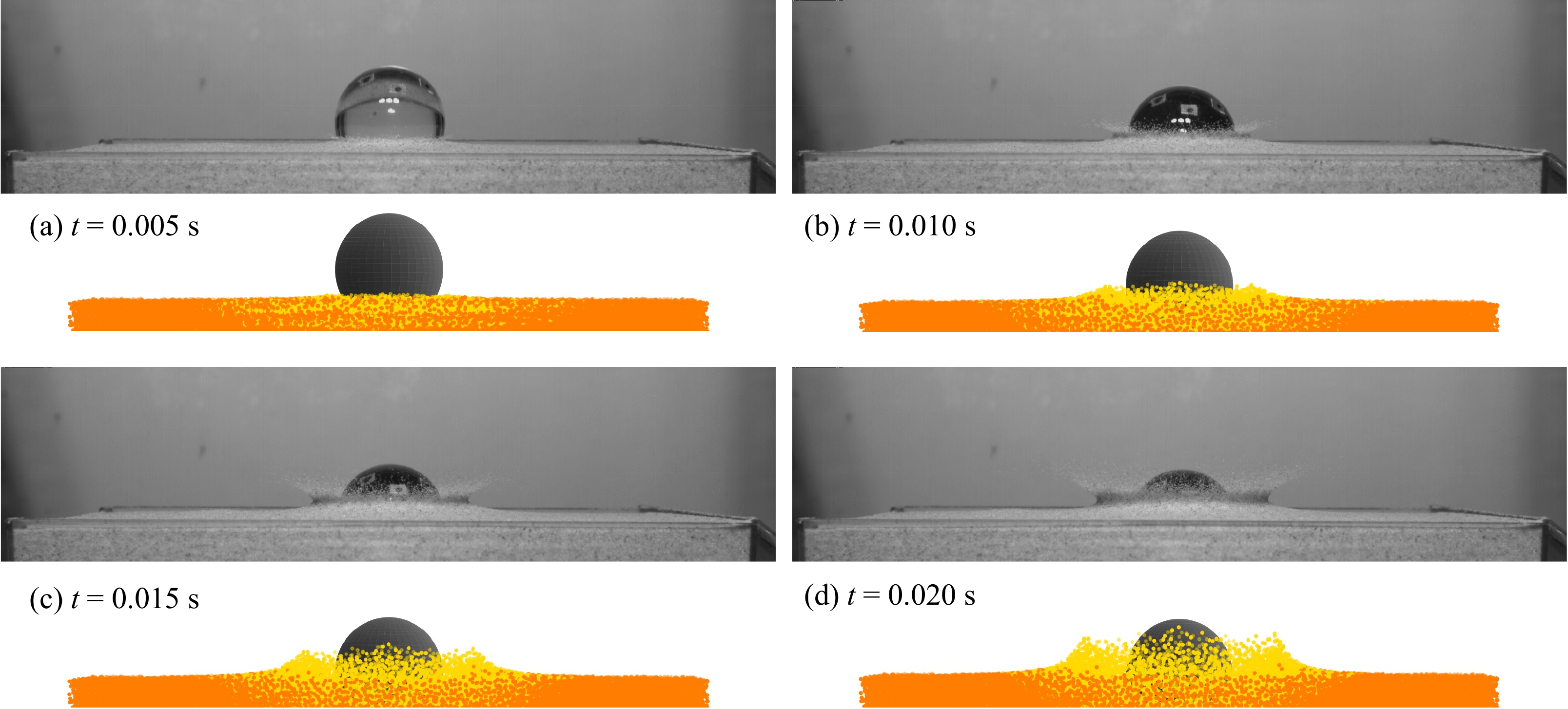}
    \caption{Comparison of splash patterns in the experiment and simulation: glass sphere dropped from $H = 40$ cm. Light-colored material points are gasified.}
    \label{fig:splash-glass}
\end{figure}

\begin{figure}[h!]
    \centering
    \includegraphics[width=1.0\textwidth]{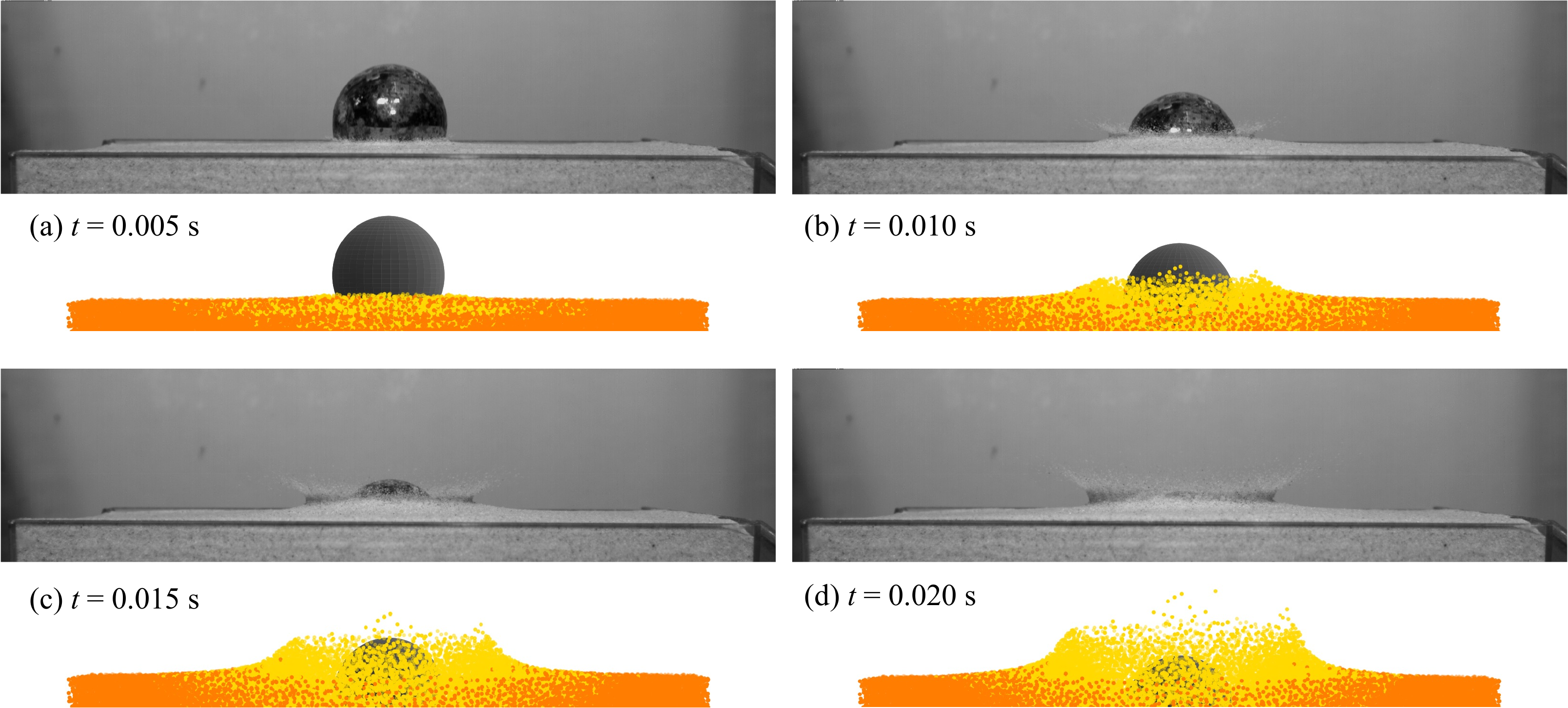}
    \caption{Comparison of splash patterns in the experiment and simulation: steel sphere dropped from $H = 40$ cm. Light-colored material points are gasified.}
    \label{fig:splash-steel}
\end{figure}

First, for qualitative validation, in Figs.~\ref{fig:splash-glass} and~\ref{fig:splash-steel} we compare the splash patterns in the experimental and simulation results of the glass and steel sphere impacts, respectively, when $H = 40$ cm.
In the simulation results, the gasified material points are distinguished by lighter colors such that they can be compared with the detached grains in the experimental images.
As can be seen, the splash patterns and their evolutions in the experiments and simulations are overall quite similar.
One may notice that the thin ``clouds'' observed in the experiments are absent in the simulation results.
This difference, however, is not surprising because a continuum representation of sand is inherently incapable of modeling separation of grains even with the trans-phase constitutive relation. 
Other than this, the splash behavior is well simulated.

\begin{figure}[h!]
    \centering
    \subfloat[Impact load]{\includegraphics[width=0.5\textwidth]{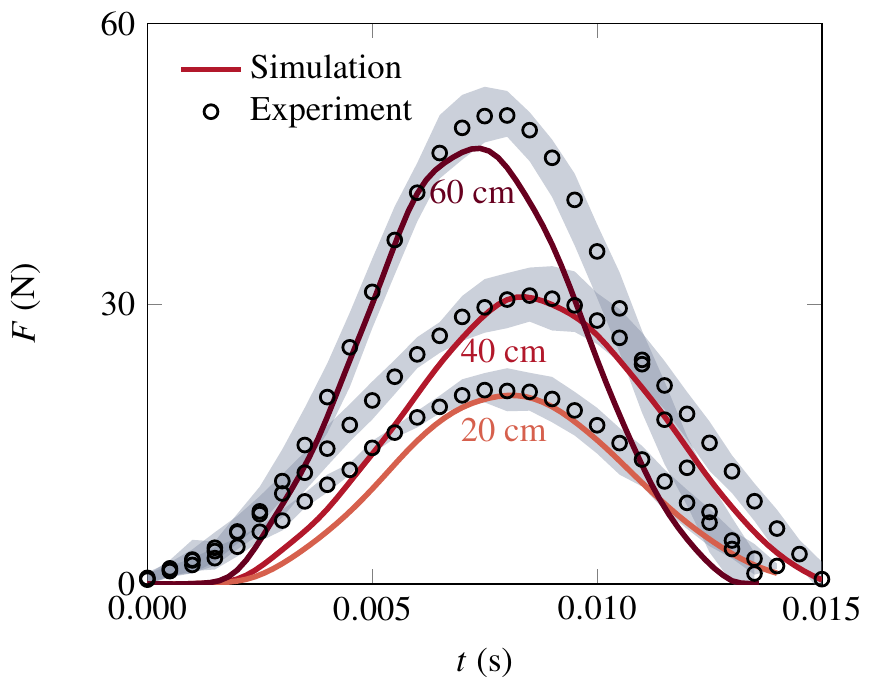}}
    \subfloat[Intrusion depth]{\includegraphics[width=0.5\textwidth]{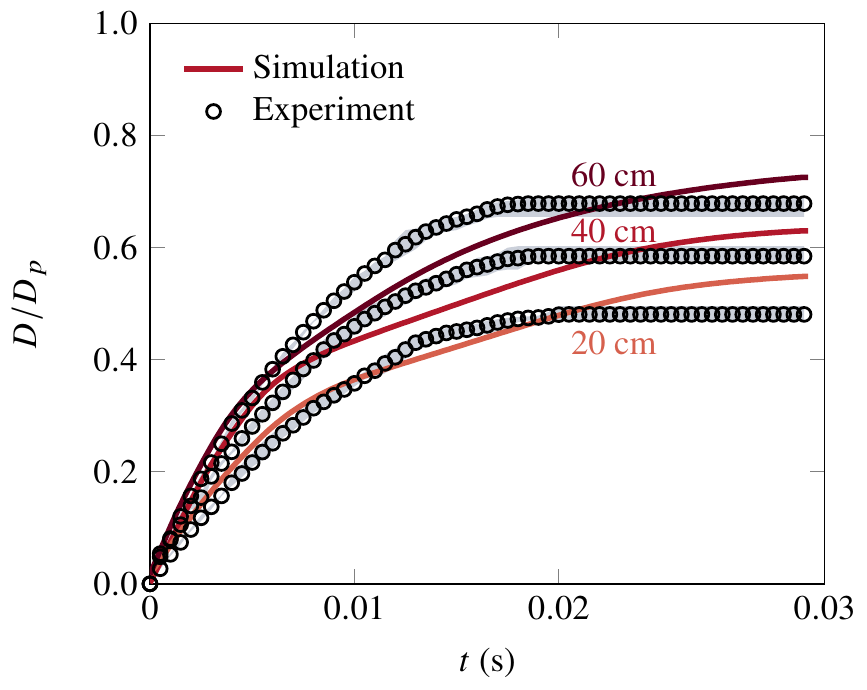}}
    \caption{Comparison of the experimental and simulation results in terms of (a) the impact load and (b) the intrusion depth (normalized by the sphere diameter): \revised{glass spheres dropped from three different heights.} \revised{The open symbols and shaded areas denote the averages and ranges, respectively, of the experimental data obtained from six repeated tests.}}
    \label{fig:validation-glass}
\end{figure}

\begin{figure}[h!]
    \centering
    \subfloat[Impact load]{\includegraphics[width=0.5\textwidth]{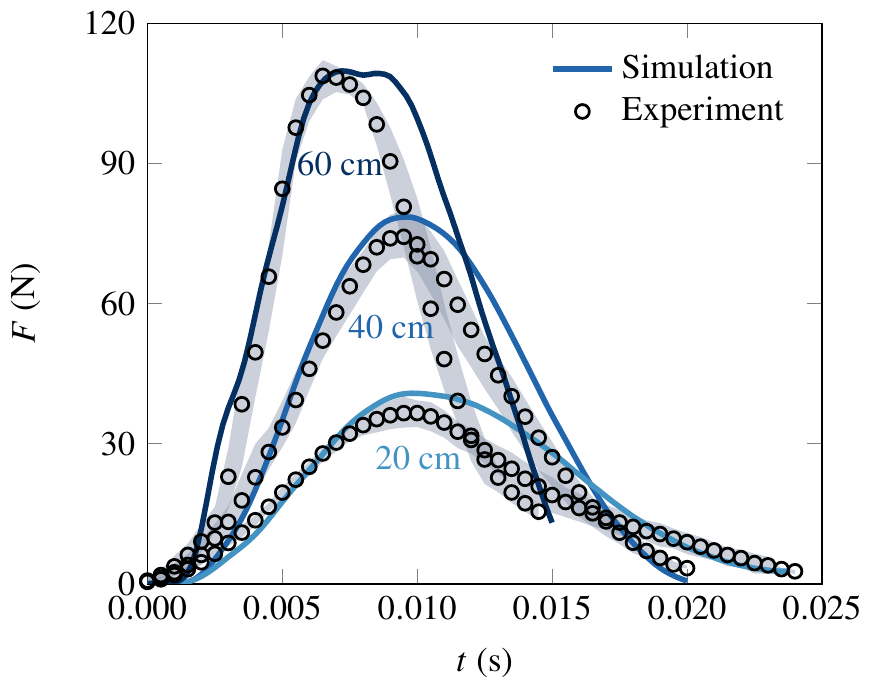}}
    \subfloat[Intrusion depth]{\includegraphics[width=0.5\textwidth]{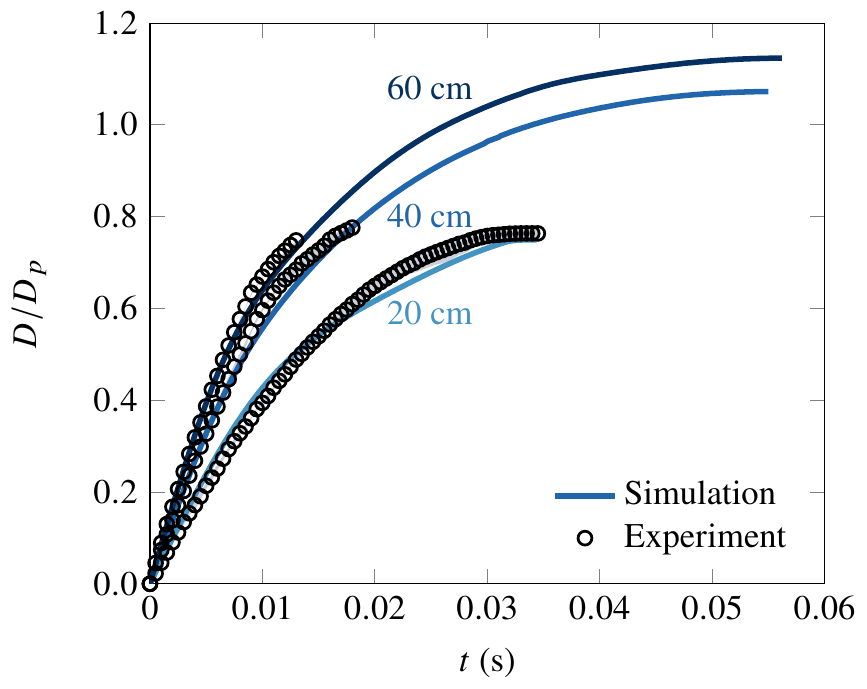}}
    \caption{Comparison of the experimental and simulation results in terms of (a) the impact load and (b) the intrusion depth (normalized by the sphere diameter): \revised{The open symbols and shaded areas denote the averages and ranges, respectively, of the experimental data obtained from six repeated tests.}}
    \label{fig:validation-steel}
\end{figure}

\begin{figure}[h!]
    \centering
    \includegraphics[width=0.5\textwidth]{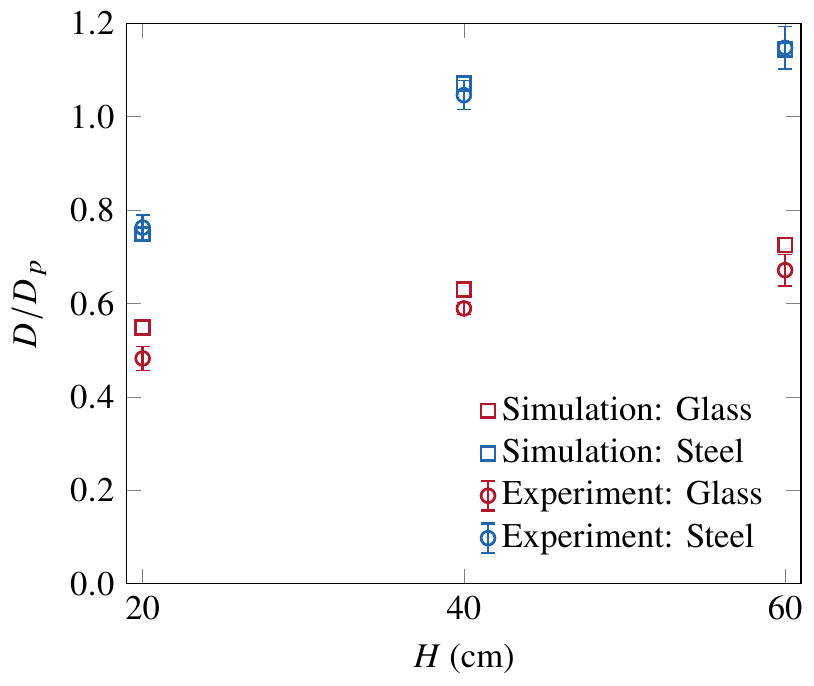}
    \caption{\revised{Comparison of the experimental and simulation results in terms of the final depth (normalized by the sphere diameter). The open symbols and error bars denote the averages and ranges, respectively, of the experimental data obtained from six repeated tests.}}
    \label{fig:validation-final-depth}
\end{figure}

Next, for quantitative validation, in Figs.~\ref{fig:validation-glass}--\ref{fig:validation-final-depth} we compare the experimental and simulation results of the impact loads, intrusion depths, \revised{and the final depths} of the glass and steel spheres.
\revised{We note that when the steel sphere was dropped from $H = 40$ cm and $60$ cm, it was impossible to estimate the intrusion depth until about $t=0.02$ s and $t=0.01$ s, respectively, because after that the sphere becomes invisible as shown in Fig.~\ref{fig:splash-steel}(d).
As shown in the figures, the experimental results on the impact load, the intrusion depth, and the final depth are highly dependent the type of solid intruder material and the dropping height.
Despite this dependence, all the experimental results are well simulated by the proposed approach with a single set of parameters.}
Particularly, the peak impact loads and final intrusion depths, which are of primary interest in practice, are well estimated for both the glass and steel sphere impacts \revised{from different dropping heights}.

The foregoing comparisons have validated that the proposed hybrid continuum--discrete approach can well reproduce granular impact dynamics, in both qualitative and quantitative manners. 
We note that the hybrid approach not only requires a fraction of the computational cost of fully discrete modeling but it also is more efficient than a fully continuum approach that models the solid intruder as a continuum.

\section{Parameter studies}
\label{sec:paramter-studies}
In this section, we conduct parameter studies to identify key factors for successful continuum--discrete simulation of granular impact dynamics.
The parameters studied are as follows: Young's modulus, friction angle, rate-dependent friction, gasification, and the MPM scheme.
For brevity, we shall focus on the glass sphere \revised{dropped from $H=40$ cm}.

\subsection{Effects of Young's modulus}
\begin{figure}[h!]
    \centering
    \subfloat[Impact load]{\includegraphics[width=0.5\textwidth]{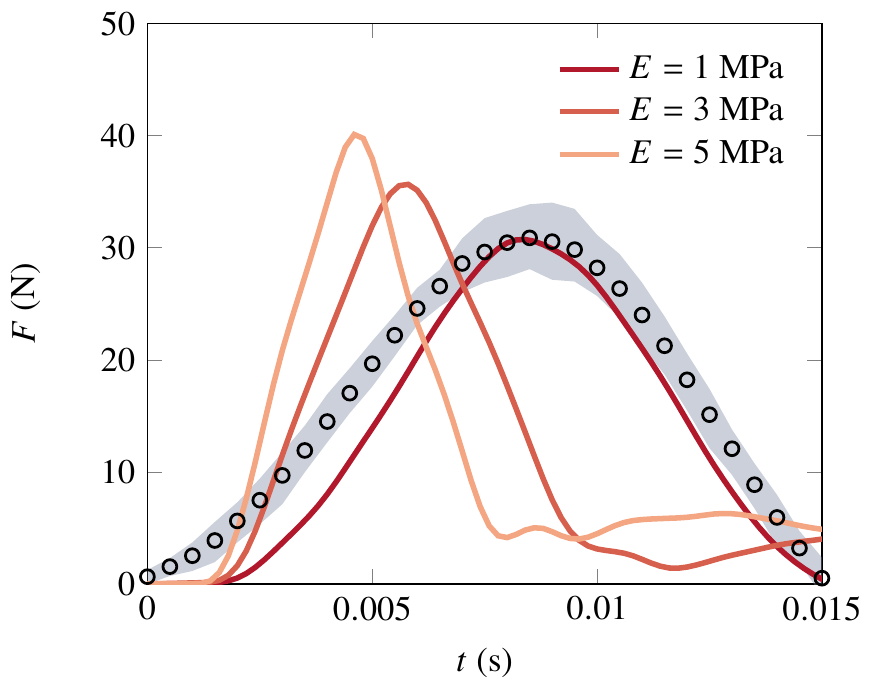}}
    \subfloat[Intrusion depth]{\includegraphics[width=0.5\textwidth]{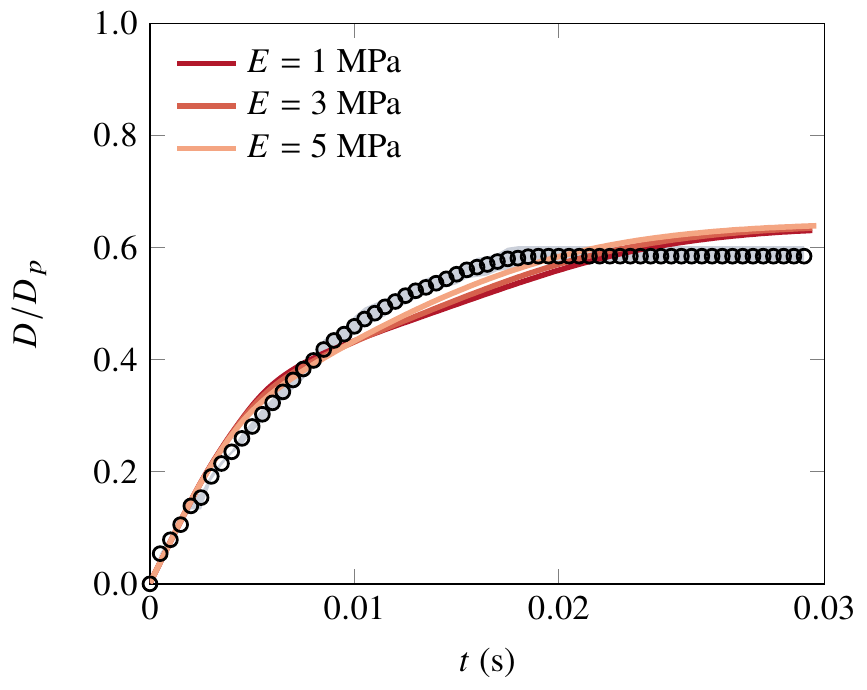}}
    \caption{Effects of Young's modulus, $E$, on (a) the impact load and (b) the intrusion depth (normalized by the sphere diameter). \revised{The open symbols and shaded areas denote the averages and ranges, respectively, of the experimental data obtained from six repeated tests.}}
    \label{fig:youngs-effects}
\end{figure}

Figure~\ref{fig:youngs-effects} shows how the Young's modulus ($E$) of sand controls the simulation results of the impact load and the intrusion depth.
As Young's modulus increases, the impact load rises faster and exhibits a higher peak value, becoming analogous to a non-adhesive pure elastic impact~\cite{bourrier2011multi,ng2021impact}.
One can also see that the double-peak pattern in the 1 MPa case disappears when Young's modulus is higher.
Conversely, the intrusion depth is virtually unaffected by Young's modulus.
This difference indicates that while the elasticity of granular media has significant effects on both the time evolution and the magnitude of the impact load, it does not exert control over the kinematics of solid intrusion.
Therefore, the elastic stiffness is important for an accurate estimation of the impact load but not for the intrusion profile.

\subsection{Effects of friction angle}
\begin{figure}[h!]
    \centering
    \subfloat[Impact load]{\includegraphics[width=0.5\textwidth]{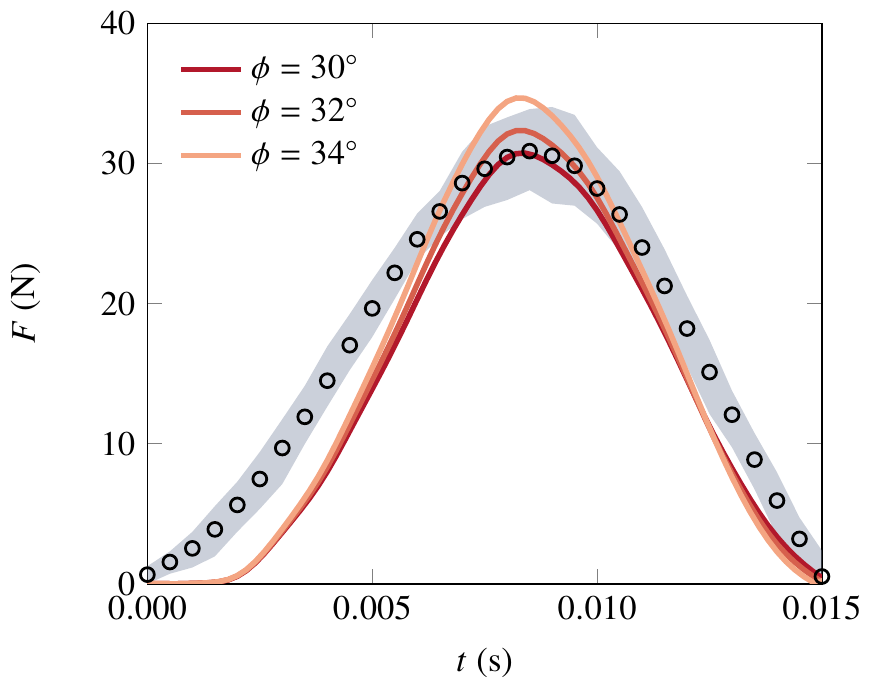}}
    \subfloat[Intrusion depth]{\includegraphics[width=0.5\textwidth]{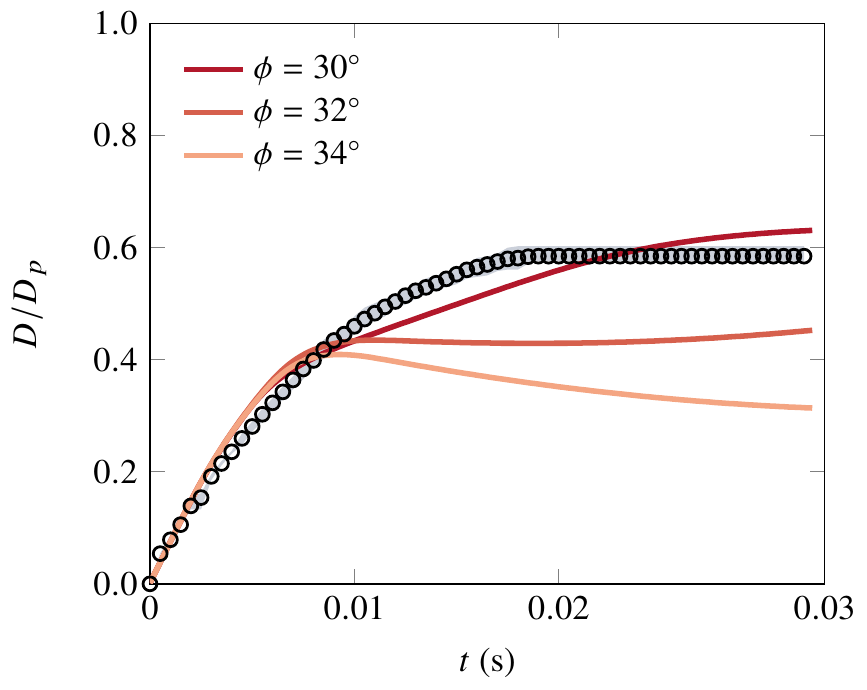}}
    \caption{Effects of friction angle, $\phi$, on (a) the impact load and (b) the intrusion depth (normalized by the sphere diameter). \revised{The open symbols and shaded areas denote the averages and ranges, respectively, of the experimental data obtained from six repeated tests.}}
    \label{fig:friction-angle-effects}
\end{figure}

In Fig.~\ref{fig:friction-angle-effects} we examine the effects of the friction angle ($\phi$) on the impact load and the intrusion depth.
One can see that the friction angle, which controls the yield strength, is a main factor for both the force and kinematics of granular impact.  
An increase in the friction angle makes the impact load shows a higher peak value, although it has a marginal effect on the time evolution of the impact load.
Also, as the friction angle becomes higher, the intrusion depth decreases.
This is because a smaller region of sand undergoes plastic deformation as the friction angle increases.

\subsection{Effects of rate-dependent friction}
To investigate whether the rate dependence of friction plays an important role in the granular impact problem at hand, we repeat the simulation using the rate-dependent $\mu(I)$ rheology model.
The lower and upper limits of the frictional resistance in the $\mu(I)$ model ($\mu_s$ and $\mu_2$) are set to be equivalent to the lowest and highest friction angles ($30^{\circ}$ and $34^{\circ}$) in the Drucker--Prager results (see Fig.~\ref{fig:friction-angle-effects}).
The remaining parameters of the $\mu(I)$ model are assigned as $I_0$ = 0.278 and $d$ = 0.25 mm based on the literature~\cite{moriguchi2009estimating, dunatunga2017continuum}.

\begin{figure}[h!]
    \centering
    \subfloat[Impact load]{\includegraphics[width=0.5\textwidth]{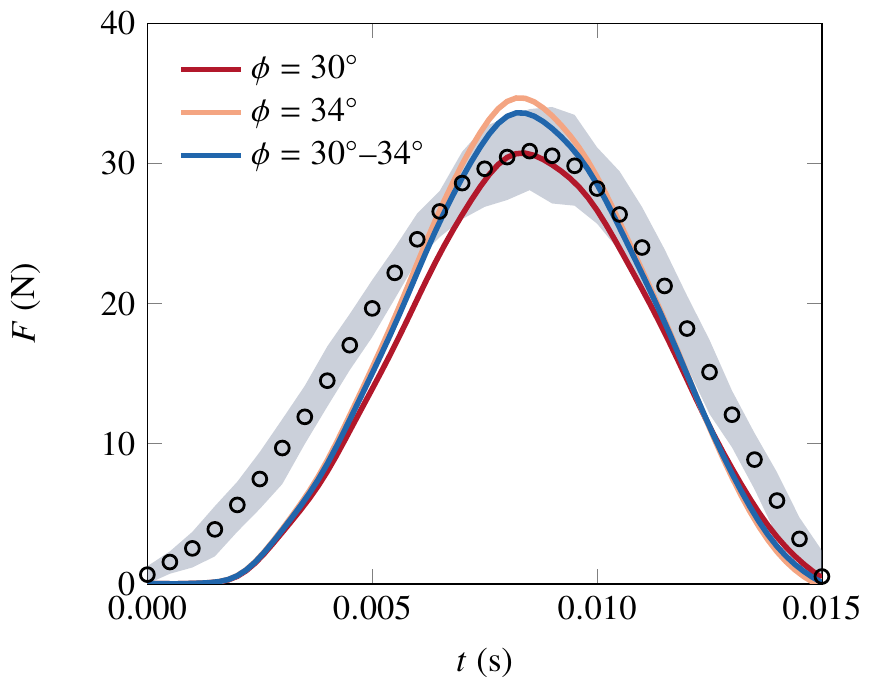}}
    \subfloat[Intrusion depth]{\includegraphics[width=0.5\textwidth]{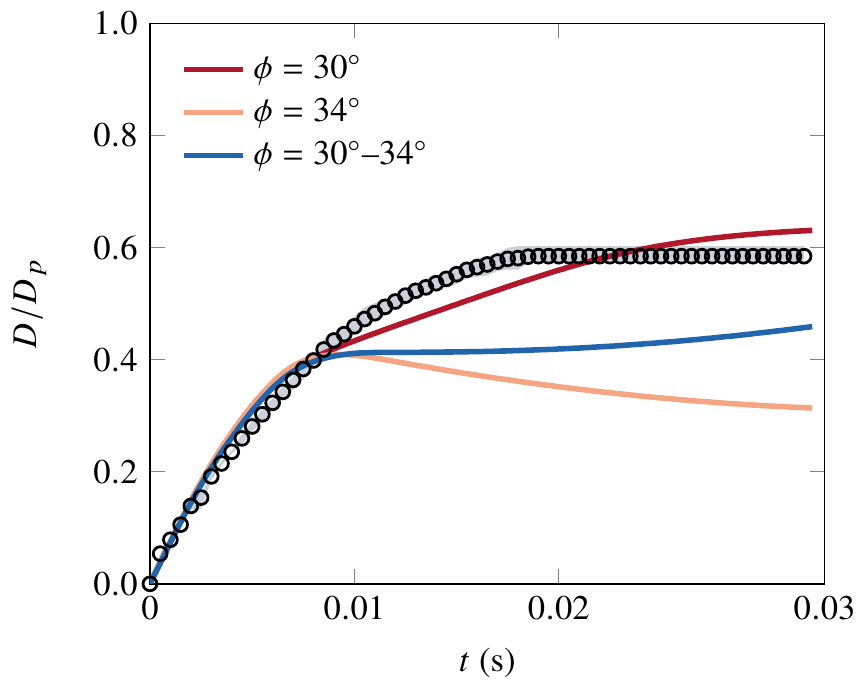}}
    \caption{Effects of the rate dependence of friction angle on (a) the impact load and (b) the intrusion depth (normalized by the sphere diameter). The case of $\phi=30^{\circ}$--$34^{\circ}$ is produced by the rate-dependent $\mu(I)$ model, whereas the other cases are produced by the rate-independent Drucker--Prager model. \revised{The open symbols and shaded areas denote the averages and ranges, respectively, of the experimental data obtained from six repeated tests.}}
    \label{fig:mu-I-effects}
\end{figure}

Figure~\ref{fig:mu-I-effects} compares the simulation results obtained with rate-dependent ($\mu(I)$) and independent (Drucker--Prager) friction angles, in terms of the impact load and the intrusion depth. We find that the results obtained with the rate-dependent friction angle lie in between those with the lower and upper limits of the friction angle.
This observation is in agreement with the continuum simulation results of Dunatunga and Kamrin~\cite{dunatunga2017continuum} which show little difference when the Drucker--Prager model is used in lieu of the $\mu(I)$ model.
Thus it would be reasonably good to use rate-independent plasticity like the Drucker--Prager model for this kind of shallow granular impact.
It is noted, however, that the rate dependence of friction might become more important as the solid intruder penetrates more deeply and hence involves a wider region of plastic deformation.

\subsection{Effects of gasification}
\begin{figure}[h!]
    \centering
    \subfloat[Impact load]{\includegraphics[width = 0.5\textwidth]{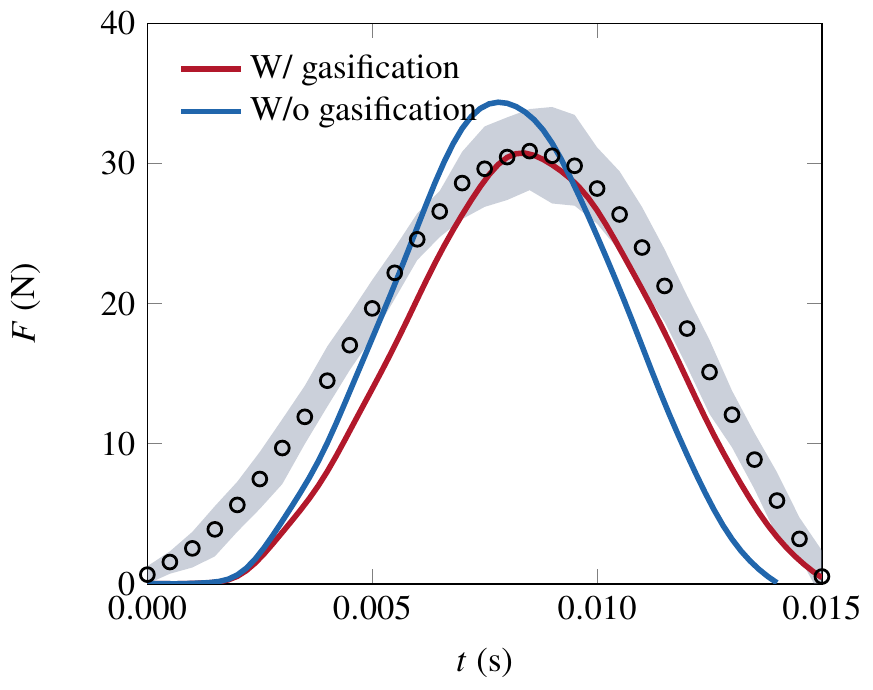}}\\
    \subfloat[Splash]{\includegraphics[width = 0.7\textwidth]{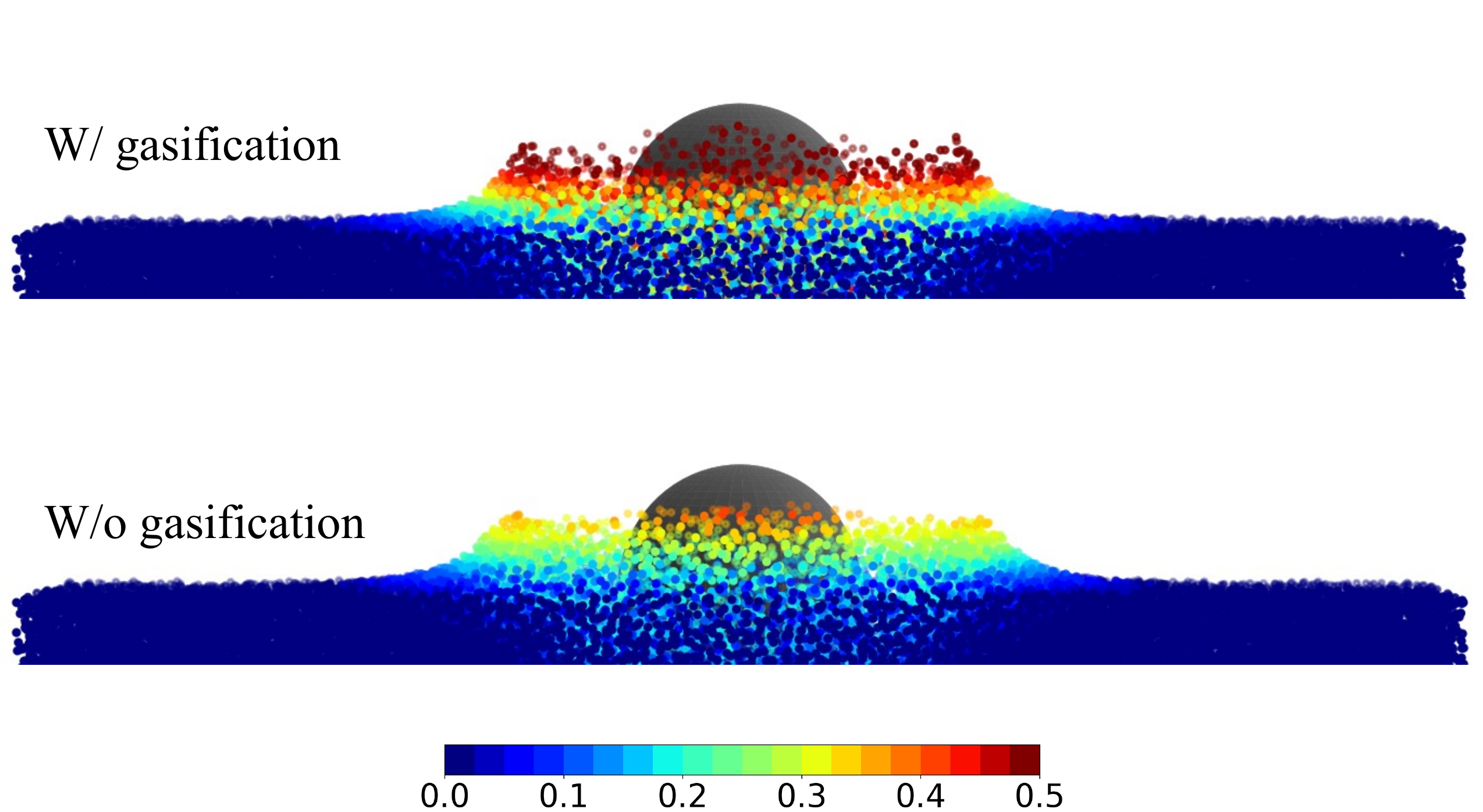}}
    \caption{Comparison of (a) the impact load and (b) splash patterns simulated with and without gasification. \revised{The open symbols and shaded areas denote the averages and ranges, respectively, of the experimental data obtained from six repeated tests.} The color bar denotes the magnitude of $v_z$ in m/s.}
    \label{fig:gasification-effects}
\end{figure}

To examine the effects of gasification, we repeat the simulation without using the trans-phase constitutive relation and compare the results with the default ones in Fig.~\ref{fig:gasification-effects}.
It can be seen that when gasification is ignored, the impact load becomes higher and the splash zone becomes wider. 
It is also found that the double-peak pattern becomes amplified in the absence of gasification.
The results thus indicate that if gasification is not considered, granular media would be simulated in a stiffer and less separable manner than they actually are.

\subsection{Effects of MPM scheme}
\begin{figure}[h!]
    \centering
    \subfloat[Impact load]{\includegraphics[width=0.5\textwidth]{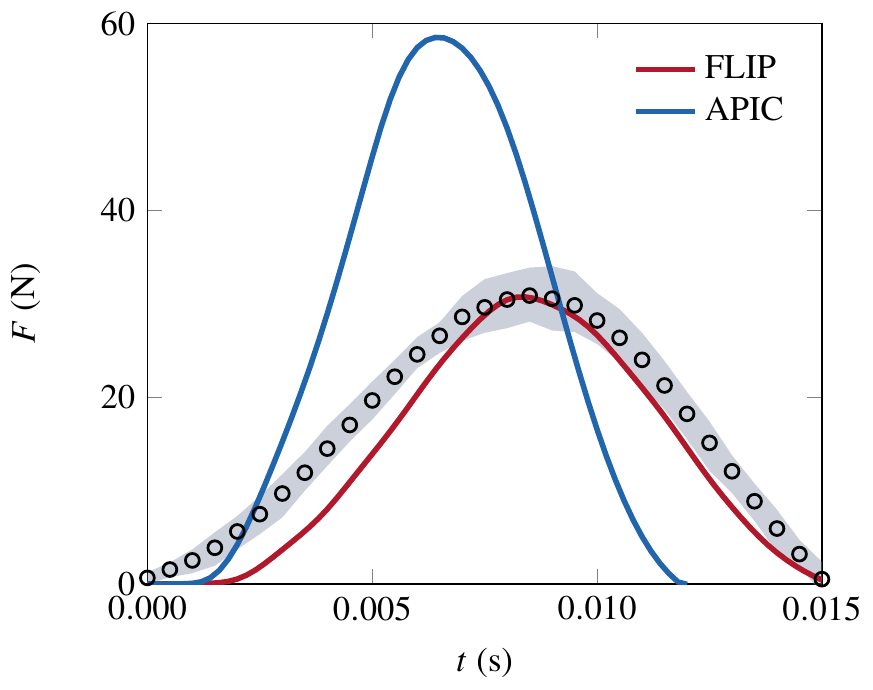}}\\
    \subfloat[Splash]{\includegraphics[width=0.7\textwidth]{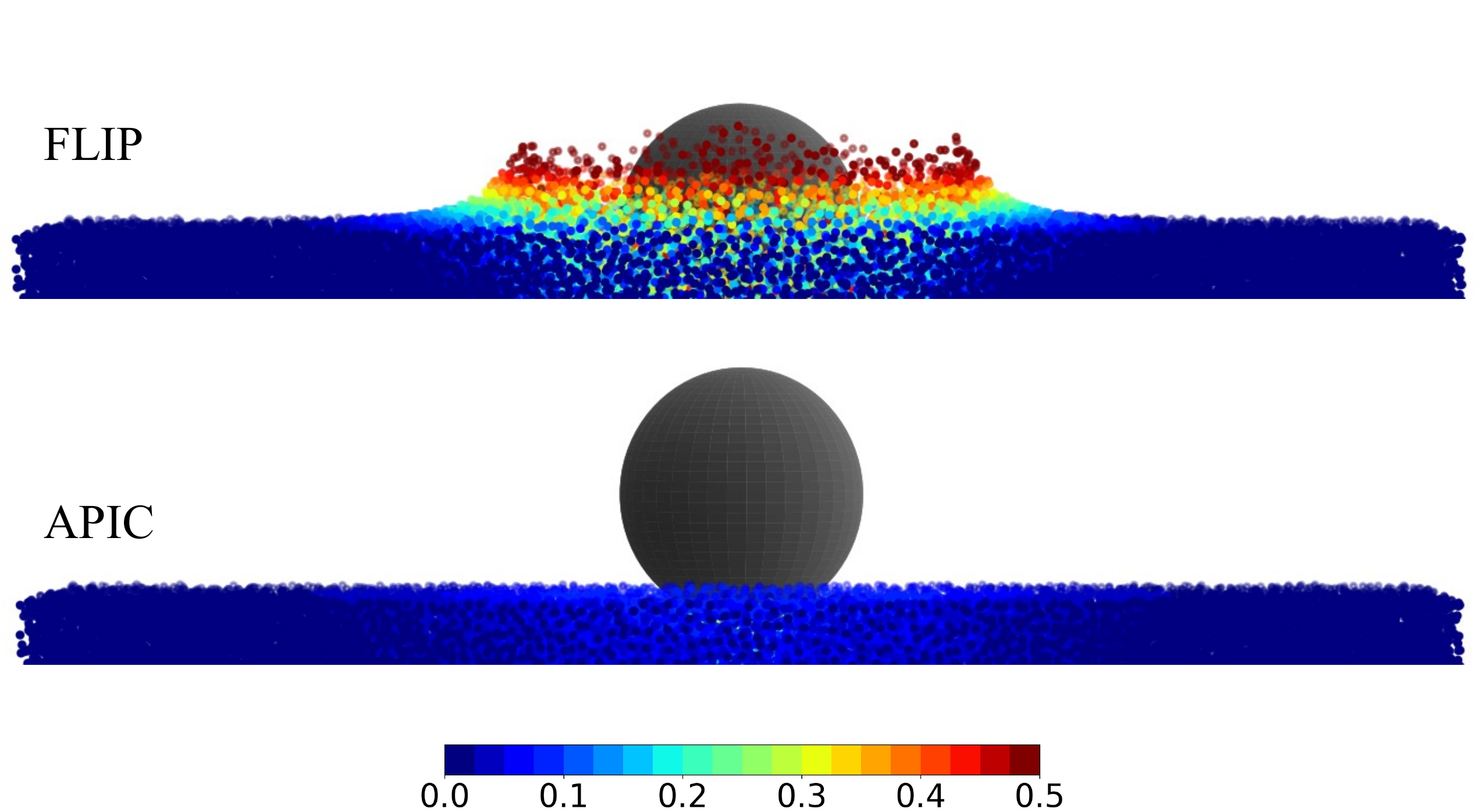}}
    \caption{Comparison of (a) the impact load and (b) splash patterns simulated with FLIP and APIC. \revised{The open symbols and shaded areas denote the averages and ranges, respectively, of the experimental data obtained from six repeated tests.} The color bar denotes the magnitude of $v_z$ in m/s.}
    \label{fig:flip-apic-effects}
\end{figure}

Lastly, we investigate the effect of the MPM scheme by repeating the simulation using APIC and compare the FLIP (default) and APIC results in Fig.~\ref{fig:flip-apic-effects}. 
When APIC is used, the simulated impact exhibits higher loads within a shorter period of time. 
Also, the sand is much less penetrated by the solid sphere without any significant splash.
The results indicate that the choice of the MPM scheme is critical to accurate simulation of granular impact dynamics. 
Specifically, the PIC update scheme, which uses total velocity values stored in the grid, significantly reduces local variations in the velocities of material points.
While this characteristic is conducive to numerical stability, it gives rise to excessive numerical damping and makes material points not sufficiently separable. 
On the other hand, the FLIP update scheme uses incremental velocity values and thus leads to less numerical damping and better separability of material points. 
As such, FLIP-based MPM is more recommended for the simulation of granular impact.

\section{Closure}
\label{closure}

We have presented a hybrid continuum--discrete approach to the simulation of granular impact dynamics.
The approach has enhanced the existing MP--DEM formulation to accommodate highly complex solid--granular interactions, in both theoretical and algorithmic aspects.
One remarkable enhancement is achieved by devising a barrier method that rigorously couples a material point and a discrete element without any interpenetration under high impact load.
Through laboratory experiments and their simulation, we have validated that the proposed approach can well reproduce the dynamics of granular impact in both quantitative and qualitative manners.
A series of parameter studies have also been conducted to clarify how material parameters and modeling approaches control different aspects of simulation results, namely, the impact load, the intrusion depth, and the splash pattern.

The hybrid continuum--discrete approach is believed to be one of the most attractive means for simulation of a variety of granular impact problems across science and engineering.
Continuum modeling, if properly formulated, can simulate the dynamics of granular media with orders of magnitude less computational cost than its fully discrete counterpart.
Another appealing aspect of continuum modeling is that it allows one to simulate a wide range of materials -- beyond dry, clean granular materials -- by taking advantage of various existing material models formulated at finite strains (\eg~\cite{borja1998cam,borja2006critical,borja2016cam,choo2018coupled,oliynyk2021finite}).
These features are highly desired for a large number of practical problems for which it is virtually impossible to model granular media in a discrete way. 
For these reasons, the hybrid continuum--discrete approach provides opportunities to better understand and predict complex granular impact dynamics in many scientific and engineering applications including geotechnics.

\section*{Author Contributions} 
\label{sec:credit}

\textbf{Yupeng Jiang}: Conceptualization, Methodology, Software, Validation, Formal Analysis, Investigation, Data Curation, Writing - Original Draft, Visualization.
\textbf{Yidong Zhao}: Methodology, Software, Investigation.
\textbf{Clarence E. Choi}: Conceptualization, Methodology, Validation, Resources, Writing - Review \& Editing, Supervision.
\textbf{Jinhyun Choo}: Conceptualization, Methodology, Validation, Investigation, Writing - Original Draft, Writing - Review \& Editing, Visualization, Supervision, Project Administration, Funding Acquisition.

\section*{Acknowledgments}
The authors are grateful to Jianting Du, Pengjia Song, and Timothy Xiong for their help with the laboratory experiment.
\revised{They also wish to thank the anonymous reviewer for the comments that helped improve the quality of the paper significantly.}
Portions of the work were supported by the Research Grants Council of Hong Kong through Projects 17201419 and 16212618. 
YZ and JC also acknowledge financial support from KAIST.

\section*{Data Availability Statement} 
\label{sec:data-availability} 

The data that support the findings of this study are available from the corresponding author upon reasonable request.

\bibliography{references}

\end{document}